\documentclass[12pt]{article}

\usepackage{newtxtext,newtxmath, bm}
\usepackage{graphicx}

\usepackage[letterpaper,margin=1in]{geometry}

\renewenvironment{abstract}
	{\quotation}
	{\endquotation}

\date{}

\makeatletter
\renewcommand{\fnum@figure}{\textbf{Figure \thefigure}}
\renewcommand{\fnum@table}{\textbf{Table \thetable}}
\makeatother

\usepackage{scicite}

\usepackage{url}

\def\scititle{
	Broadband uniform-efficiency OAM-mode detector
}

\title{\bfseries \boldmath \scititle}

\author{
	Suman Karan$^{1\ast}$,
	Martin P. Van Exter$^{2}$,
	and Anand K. Jha$^{1 \ast}$\and
	\small$^{1}$Department of Physics, Indian Institute of Technology Kanpur, Kanpur, UP 208016, India.\and
	\small$^{2}$Huygens-Kamerlingh Onnes Laboratory, Leiden University, P.O.Box 9504, 2300 RA Leiden,The Netherlands.\and
	\small$^\ast$Corresponding author. Email: karans@iitk.ac.in, akjha@iitk.ac.in.
}

\begin{document} 

% Insert the title and author list
\maketitle

% Abstract, in bold
% There are strict length limits, and not all formats have abstracts.
% Consult the journal instructions to authors for details.
% Do not cite any references in the abstract.
\begin{abstract} \bfseries \boldmath
The high-dimensional basis of orbital angular momentum (OAM) has several added and unique advantages for photonics quantum technologies compared to the polarization basis, which is only two-dimensional. However, one of the major roadblocks in implementing OAM-based applications with their full potentials is the absence of an ideal OAM-mode detector. Despite the plethora of efforts in the last three decades, currently, there is no OAM detector that can detect a broad OAM-mode spectrum, has uniform detection-efficiency over all the modes, measures the true spectrum, and works for an arbitrary quantum state without the need for any prior information. In this article, we  experimentally demonstrate just such an OAM detector. We report detection of pure and  mixed OAM states with fidelities more than 98$\%$ and with measurement times of only a few minutes for dimensionalities up to 100. We expect our work to substantially  boost the OAM-based photonics quantum technology efforts.

\end{abstract}

% The first paragraph of any Science paper does NOT have a heading
% Nor is it indented
\noindent
%%%%%%%%%%%%%%%%%%%%%
\section*{\label{sec:intro} Introduction}
In 1992, Allen {\it et al.} showed that a photon in a Laguerre-Gaussian (LG) mode, represented as $LG_p^{|l|}(\rho)e^{il\phi}$ in the transverse polar coordinates, carries $l\hbar$ orbital angular momentum (OAM). Here $l$ is an integer ranging from $-\infty$ to $\infty$ and is referred to as the OAM-mode index whereas $p$ is called the radial-mode index, ranges from $0$ to $\infty$, and decides the radial profile of the mode \cite{allen1992pra}. As the OAM of a photon provides a basis that is not only high-dimensional but also discrete, several OAM-based applications have been proposed and demonstrated, highlighting the added and unique advantages of the high-dimensional OAM basis for photonic quantum technologies compared to the two-dimensional polarization basis \cite{willner2015aop, erhard2018lsa}. These include demonstrating terabit-scale data transmission using OAM-mode multiplexing for long distance communication through free-space \cite{wang2012natphoton, yan2014natcomm} and fiber \cite{bozinovic2013science}, enhanced security and error tolerance for quantum communication protocols \cite{cerf2002prl,nikolopoulos2006pra}, efficient gate implementation \cite{ralph2007pra,lanyon2009naturephys}, super-sensitive measurement in quantum metrology \cite{dambrosio2013natcomm, jha2011pra}, and fundamental tests of quantum mechanics \cite{kaszlikowski2000prl,collins2002prl,vertesi2010prl,dada2011naturephys}.  The OAM basis has been shown to be the preferred basis for ground-to-ground and ground-to-satellite long-distance communications\cite{dambrosio2012natcomm,vallone2014prl, bhattacharjee2022sciadv}. More recently, the OAM basis has been used for demonstrating quantum cryptography within a city \cite{sit2017optica} and through outdoor underwater channel \cite{bouchard2018optexp}. High-dimensional quantum gates \cite{brandt2020optica} and entanglement distribution \cite{hu2020optica} have been demonstrated with OAM modes. As an important advancement for long-distance communication, optical fibers have been demonstrated that can carry several OAM modes upto a kilometer without  substantial loss or cross-talk \cite{ma2023science}.

Although the push for exploiting OAM basis for quantum information technologies has been steadily increasing in the past three decades, one of the major roadblocks for implementing OAM-based applications with their full potential is the absence of an OAM detector that detects a broad range of modes with uniform detection efficiency, measures the true spectrum, and works for an arbitrary quantum state without the need for any prior information about the state. Consequently, developing such a detector has been an active area of research \cite{mair2001nature,heckenberg1992optlett,pires2010optlett, pires2010prl,jha2011pra,malik2012pra,vasnetsov2003optlett,zhou2017lsa,leach2002prl}.  
There are several approaches to detecting the OAM of photonic quantum states. 

One main approach \cite{mair2001nature, heckenberg1992optlett}, that is currently the most widely used, is based on displaying holograms specific to different OAM modes onto a spatial light modulator (SLM) and then measuring the intensity of the $(l=0$, $p=0)$ mode at the
first diffraction order of the SLM using a single-mode fiber (SMF). This way, by placing holograms specific to different input OAM-modes in a sequential manner, one is able to measure the spectrum. However, an SMF has non-uniform coupling efficiencies for different OAM-modes \cite{qassim2014josab} and therefore the efficiency of this method decreases with increasing $l$, making it unsuitable for broadband detection. Furthermore, an SMF detects primarily the $p=0$ mode and therefore this method does not work for quantum states in which the OAM is distributed over multiple radial modes \cite{torres2003pra, kulkarni2017natcomm}. Even the p=0 mode detection can be efficient only if one has the prior knowledge of the generation beam waist. (See Supplementary Materials section 1 for a detailed analysis.)  The SMF based OAM detection method has also been integrated with compressive sensing techniques for faster measurement of up to 17-dimensional nearly-pure states in the OAM basis \cite{tonolini2014scirep}. However, besides the fact that this method works only for nearly-pure states and requires prior knowledge of the state in order to achieve faster detection, it suffers from all the limitations faced by the SMF based schemes. More recently, the traditional SMF-based techniques for measuring just the $p=0$ modes, which works on the principle of phase flattening, has been extended to also detect $p\neq 0$ radial modes  by implementing a smart technique referred to as the intensity-flattening \cite{bouchard2018optexp2}. With this, detection of up to 8 radial modes as well as the first 55 lowest order LG modes have been demonstrated. However, this detector suffers from considerable loss, and requires prior information about the mode-space and its dimensionality in order to optimize the detection efficiency of different modes. Even more recently, advanced techniques for measuring spatial modes have been demonstrated using multiple SLMs and an SMF \cite{hiekkamaeki2019oe, shahar2024aplphoton, choudhary2018ol}. These methods provide more control as far as optimizing the mode-dependent efficiency is concerned. Nonetheless, due to the inherent reflections at multiple SLMs, the loss in general is high, and for the better optimization of detection efficiency, one need to have prior information about the input state.

\begin{figure*}[t!]
\centering
\includegraphics[scale=0.94]{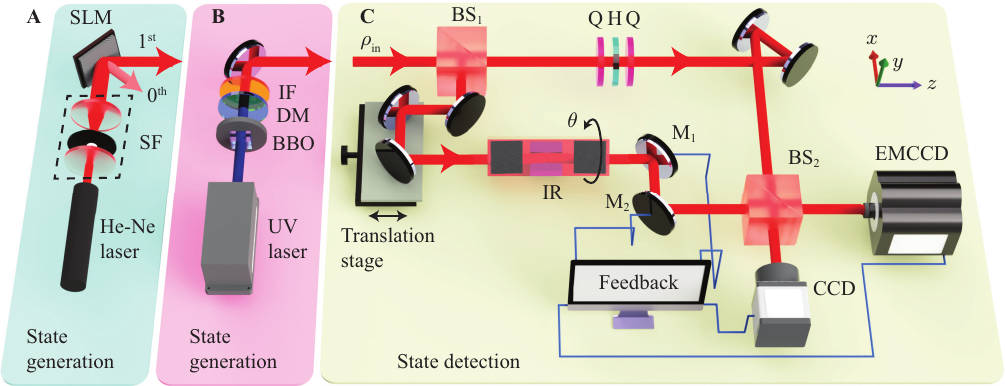}
\caption{{\bf Experimental setup.} {\bf(A)} Schematic of the experimental setup for generating single-photon quantum states using an attenuated He-Ne laser and an SLM. {\bf(B)} Schematic of the experimental setup for generating single-photon quantum states through SPDC.  {\bf(C)} Schematic of the experimental setup of the proposed OAM detector. Q and H stand for quarter-and half-wave plates. The combination Q-H-Q is geometric phase unit comprising a quarter-wave plate (Q)  at $45^{\circ}$, a half-wave plate (H) at $\delta/2$, and another quarter-wave plate  at $45^{\circ}$, resulting in a geometric phase of $ \delta$.  The angular deviation of the beam caused by IR is adjusted by two electronically controlled mirrors $M_1$ and $M_2$ using an automatic feedback mechanism. CCD is the charged coupled device camera and EMCCD is an electron-multiplied CCD. In the above figure, DM is the dichroic mirror to block the UV pump; IF is the interference filter of central wavelength $810$nm with bandwidth of $10$nm;  BS stands for a beam splitter; IR is a home-built image rotator; and SF is a spatial filter.}
\label{fig_exp_setup}
\end{figure*}

The second set of approaches, which do not use an SMF and thus do not suffer from the coupling-efficiency issues, rely on reconstructing the OAM spectrum through measuring the angular coherence function  \cite{pires2010optlett, pires2010prl, jha2011pra, malik2012pra}. An effective way to measure the angular coherence function is by measuring the visibility in a Mach-Zehnder interferometer of the interference between two optical fields that are rotated with respect to each other \cite{pires2010optlett, pires2010prl}. The experimental implementation of such OAM detectors have been carried out for diagonal mixed states at high-light levels \cite{pires2010optlett} as well as for entangled two-photon states \cite{pires2010prl}, in which case the method even becomes phase-insensitive. However, we note that the scheme in Ref.~\cite{pires2010prl} works only for measuring the OAM Schmidt spectrum of entangled photons but does not work for single photon OAM states. Another way of measuring the angular coherence function \cite{jha2011pra, malik2012pra} is by using angular double-slits \cite{jha2010prl}; however, since only a very small portion of the incident field can be utilized for detection, this method is not suitable at single-photon levels. The other approaches to measuring the OAM spectrum include techniques based on
rotational Doppler frequency shift \cite{vasnetsov2003optlett,
zhou2017lsa} and concatenated Mach-Zehnder interferometers
\cite{leach2002prl}. However, due to experimental
challenges, these approaches \cite{vasnetsov2003optlett,
zhou2017lsa, leach2002prl} have been demonstrated only for quantum states consisting of just a few OAM modes.

More recently, there have been several works that have also overcome some of the limitations mentioned above. For example, detection of 27-dimensional pure states has been demonstrated using weak measurement techniques \cite{malik2014natcomm}. Such weak measurement based technique can work for measuring a general density matrix as well, as proposed theoretically \cite{lundeen2012prl} and demonstrated experimentally \cite{zhou2021prl} in the position basis. However, there has been no experimental demonstration of weak measurement based technique for measuring an arbitrary quantum state in the OAM basis. Next, using interferometric schemes, detection of upto 25-dimensional pure states \cite{kulkarni2020prapplied} and more than 200-dimensional mixed diagonal states \cite{kulkarni2017natcomm, kulkarni2018pra} have been demonstrated.  However, these schemes have so far been demonstrated only for specific classes of quantum states and not for an arbitrary quantum state in the OAM basis. Another technique that overcomes several of the limitations mentioned above is based on the novel idea of measuring the azimuthal Wigner distribution function and thereby reconstructing the state in the OAM basis \cite{mirhosseini2016prl}. Although this technique in principle works for arbitrary high-dimensional states, it has so far been demonstrated only up to 7-dimensional states with fidelity up to only 90$\%$.

A slightly different set of efforts have been through making an OAM sorter which aims to spatially separate OAM modes based on their mode index $l$ \cite{mirhosseini2013natcomm, lavery2012optexp, berkhout2010prl, malik2014natcomm, fontaine2019natcomm, sahu2018optexp}. An ideal OAM sorter works as a perfect OAM detector. However, the current OAM sorter implementations involve either  multiple diffractive elements and thus suffer from losses that are very difficult to estimate \cite{mirhosseini2013natcomm, lavery2012optexp, berkhout2010prl, malik2014natcomm}, or sort the OAM modes with substantial overlap between neighbouring modes \cite{sahu2018optexp}, or need spatially separated modes at the input itself \cite{fontaine2019natcomm}. Therefore, these cannot be employed as OAM detectors. In contrast, in this paper, we experimentally demonstrate an OAM-mode detector that works for a broad range of modes, has uniform detection-efficiency over the entire range, measures the true mode spectrum, and works for an arbitrary quantum state without the need for any prior information.

\begin{figure*}[!t]
\centering
\includegraphics[scale=0.84]{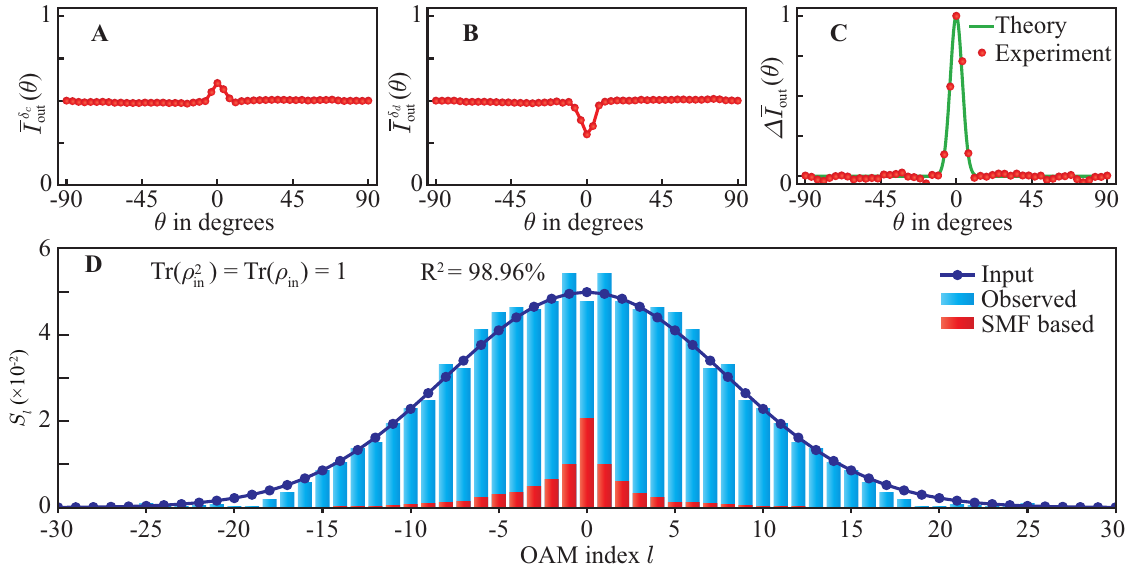}
\caption{{\bf Experimentally measured OAM spectrum for a $50$-dimensional pure state ${\rm Tr}\left(\rho^{2}_{\rm in}\right)={\rm Tr}\left(\rho_{\rm in}\right) =1$.} {\bf(A)} and {\bf(B)} are the plots of the photon detection probability $\bar{I}_{\rm out}^\delta(\theta)$ as a function of the rotation angle $\theta$ at $\delta= \delta_{c} \approx 0$ and $\delta= \delta_{d} \approx  \pi$. {\bf(C)} Experimentally measured polarization-corrected difference probability $\Delta I _{\rm out}\left( \theta \right)$ as a function of $\theta$. The red dots are the experiments and the solid green line is the theory. {\bf(D)} Reconstructed OAM spectrum. Blue bars are the measured spectrum while the blue dots are the input spectrum; the red bars represent the performance of an SMF-based detector. }
\label{fig:gaussian_pure_state_result}
\end{figure*}

\section*{\label{sec:results}Results}
\noindent {\bf Theory of OAM-spectrum detection} \\
We consider an arbitrary $2N+1$-dimensional single-photon quantum state in the OAM basis given by the density matrix:
\begin{align}\label{input_state}
\rho_{\rm in}= \sum^{+N}_{l_1,l_2=-N}\sum^{\infty}_{p_1,p_2 =0}  C^{p_1,p_2}_{l_1,l_2} |l_1, p_1\rangle\langle l_2,p_2|.
\end{align}
Here $|l_1, p_1\rangle$ is a single-photon state with mode indices ($l_1$, $p_1$) and $C^{p_1,p_2}_{l_1,l_2}$ is the density matrix element corresponding to indices $(l_1, p_1)$ and $(l_2, p_2)$. The projection of $|l_1, p_1\rangle$ on the transverse-polar basis state $|\rho,\phi\rangle$ is given by $\langle\rho,\phi|l,p\rangle = LG^{|l|}_p(\rho)e^{-il\phi}$. Figure \ref{fig_exp_setup}C shows the schematic of our experimental setup. The state $\rho_{\rm in}$ enters the interferometer having an image rotator (IR) oriented at angle $\theta$ in one of the two arms and gets detected using a single-photon sensitive electron-multiplied charge coupled device (EMCCD) camera. IR is a home-built wavefront rotating device consisting of there mirrors. As depicted in the figure, we consider the beam propagation direction to be along $\hat{\bm z}$ and the azimuthal angle $\phi =0$ to be along $\hat{\bm x}$. We define the $\theta =0$ rotation angle to be when the three mirrors are on the $x-z$ plane. The transformation of the field due to a mirror reflection can be written as $\phi \rightarrow \pi- \phi$, while that due to the IR can be written as $\phi \rightarrow -\phi+ \pi + 2\theta$. Therefore, we express the projection operators for the mirror and the IR as $\hat{M}= \sum_{l,p} e^{-i l \pi}|-l,p\rangle \langle l, p |$ and $ \hat{\rm IR}\left(\theta\right) = \sum_{l, p}  e^{-i l \left( \pi + 2\theta \right )} |-l, p\rangle \langle l, p| $, respectively (see Supplementary Materials, sections 2 and 3 for the detailed derivations). For an $\hat{\bm x}$-polarized input, the polarization state of the field after passing through the IR is given by $
\hat{\bm \epsilon}\left( \theta\right)=  \cos\left[ \psi\left(\theta\right)\right] \hat{\bm x} +  \sin\left[ \psi\left(\theta\right)\right]e^{i\chi\left(\theta\right)} \hat{\bm y}$ \cite{karan2022ao}. Therefore, the projection operator at the EMCCD plane becomes: $
\hat{P}=\sum_{l=-\infty}^{\infty}\sum_{p=0}^{\infty} e^{i(-l\pi + \beta_1 -\omega_0 t_1 + \gamma_1)}
[|k_1|\hat{\bm x} +|k_2| e^{-i\left(\delta+2 l \theta\right)}~\hat{{\bm\epsilon}}\left(\theta\right)]|-l, p\rangle\langle l, p|$, where, $\omega_0$ is the central frequency, and $k_1= |k_1|e^{i\gamma_1}$ and $k_2 = |k_2|e^{i\gamma_2}$ are the  fractions of the input field passing through the two arms; these fractions depend on the reflection and the transmission coefficients of the beam splitters and the mirrors; $\delta= (\beta_1 -\beta_2) - \omega_0(t_1 -t_2)+ \gamma_1- \gamma_2 $, where  $t_1$ and $t_2$ are the photon travel times, and $\beta_1$ and $\beta_2$ are the non-dynamical phases in the two arms. The combination Q-H-Q is utilized for introducing a geometric phase $\delta$. The density matrix at the EMCCD plane is given by $\rho_{\rm out}= \hat{P}\rho_{\rm in}\hat{P}^{\dagger}$, and thus the  total photon detection probability is given by $I^{\delta}_{\rm out}(\theta)=\int_{0}^{\infty}\int_{0}^{2\pi} \langle\rho, \phi | \rho_{\rm out} |\rho,\phi \rangle~\rho d\rho~ d\phi$. (see Supplementary Materials section 4 for detailed derivation).
\begin{align}\label{angle-av intensity}
I^{\delta}_{\rm out}(\theta)= \sum_{l=-N}^N\sum_{p=0}^{\infty} C^{p,p}_{l, l}\Big[|k_1|^2 + |k_2|^2   + 2|k_1||k_2| \cos\left[\psi\left(\theta\right)\right] \cos\left(\delta + 2 l \theta\right)\Big].
\end{align}
\begin{figure*}[t!]
\centering
\includegraphics[scale=0.84]{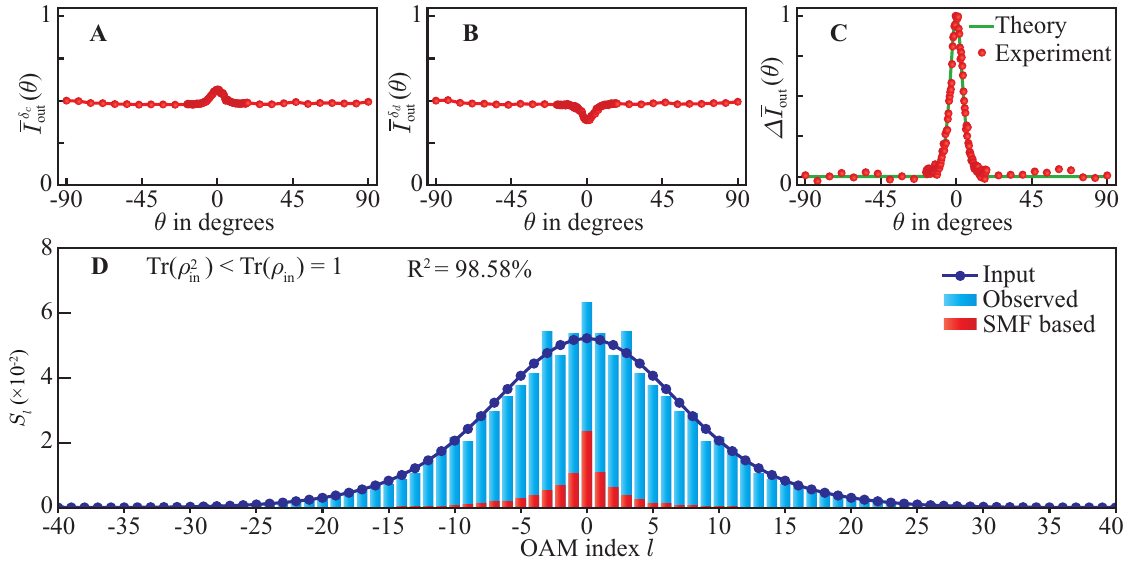}
\caption{{\bf Experimentally measured OAM probability distribution for a $80$-dimensional general mixed state ${\rm Tr}\left(\rho^{2}_{\rm in}\right) < {\rm Tr}\left(\rho_{\rm in}\right) =1$.} {\bf(A)} and {\bf(B)} are the plots of the photon detection probability $\bar{I}_{\rm out}^\delta(\theta)$ as a function of the rotation angle $\theta$ at $\delta= \delta_{c} \approx 0$ and $\delta= \delta_{d} \approx  \pi$. {\bf(C)} Experimentally measured polarization-corrected difference probability $\Delta I _{\rm out}\left( \theta \right)$ as a function of $\theta$. The red dots are the experiments and the solid green line is the theory. {\bf(D)} Reconstructed OAM spectrum. Blue bars are the measured spectrum while the blue dots are the input spectrum; the red bars represent the performance of an SMF-based detector.}
\label{fig:gaussian_gen_mixed_state_result}
\end{figure*}
The OAM spectrum of the input state $S_l$ is obtained by summing $C^{p,p}_{l,l}$ over all possible $p$ modes, i.e., $S_l=\sum^{\infty}_{p=0}C^{p,p}_{l,l}$. We note that our detection scheme does not require measurements of spatially-resolved photon detection probability and is therefore not limited by the spatial resolution of the EMCCD in contrast to several other schemes \cite{kulkarni2020prapplied, kulkarni2017natcomm, kulkarni2018pra}. In realistic experimental situations, the measured photon detection probability always contains some noise contributions $I_n^\delta(\theta)$, the sources of which in our experiments are the ambient  light and the dark count of the EMCCD. Thus the measured probability becomes $\bar{I}_{\rm out}^\delta(\theta) = I_n^\delta(\theta) + I^{\delta}_{\rm out}(\theta)$. In order to bypass such contributions, we use  the two-shot technique \cite{kulkarni2017natcomm} and perform the measurements at $\delta= \delta_{c}$ and  $\delta=\delta_{d}$. We assume that the shot-to-shot noise remains constant and that the spectrum is symmetric, that is,  $ I_n^{\delta_c}(\theta)  \approx I_n^{\delta_d}(\theta)$ and $S_l=S_{-l}$. The difference probability $\Delta\bar{I}_{\rm out}(\theta) = \bar{I}_{\rm out}^{\delta_c}(\theta) - \bar{I}_{\rm out}^{\delta_d}(\theta)$ can therefore be written as $
\Delta\bar{I}_{\rm out}(\theta)= 2|k_1||k_2|\cos[\psi\left(\theta\right)](\cos\delta_c-\cos\delta_d) \sum_{l=-N}^{+N}   S_l \cos {2 l \theta}$. For bypassing the polarization effects manifesting as $\cos\left[\psi\left(\theta\right)\right]$, we work with the polarization-corrected difference probability $\Delta I_{\rm out}(\theta)= \Delta\bar{I}_{\rm out}(\theta)/\cos\left[\psi\left(\theta\right)\right]$:   
\begin{equation}\label{deltaI_pol_corrected}
\Delta I_{\rm out}(\theta)= 2 |k_1||k_2| \sum_{l=-N}^{N}  S_l (\cos \delta_c - \cos\delta_d) \cos{2l\theta}.
\end{equation}
Since $\cos\left[\psi\left(\theta\right)\right]$ is the $\hat{\bm x}$-projection of the field, we have 
\begin{align}
\cos\left[\psi\left(\theta\right)\right]=\sqrt{\frac{|\hat{\bm \epsilon}(\theta)\cdot\hat{\bm x}|^2}{|\hat{\bm \epsilon}(\theta)|^2}}=\sqrt{\frac{I_{2x}(\theta)}{I_2^{\rm tot}}},
\end{align}
where $I_2^{\rm tot}$ and $I_{2x}(\theta)$ are the total probability and the probability along $\hat{\bm x}$ of the field passing through the IR. (See the Materials and Methods section for the experimental measurement of $\cos\left[\psi\left(\theta\right)\right]$.) We define the measured OAM spectrum $\bar{S}_l$ to be
\begin{align}
\bar{S}_l \equiv \int_0^{\pi} \Delta I_{\rm out}(\theta) \cos (2l\theta) d\theta =2 |k_1||k_2| (\cos \delta_c - \cos\delta_d) S_l.
\end{align}
We note that $\bar{S}_l$ is not normalized but is proportional to the true OAM spectrum $S_l$. Therefore, using the fact that ${\rm Tr}(\rho_{\rm in})=1$, we obtain the true OAM spectrum as:
\begin{equation}\label{eqn:normalization}
S_l = \bar{S}_l \ / \left(\sum\limits_{l=-N}^{+N} \bar{S}_l\right).
\end{equation}
To estimate the measurement accuracy of our experimental scheme, we use the coefficient of determination $R^2$:
\begin{equation}
R^2 \equiv \dfrac{\sum^{+N}_{l=-N}\left( S^{\rm ob}_l - S^{\rm in}_l\right)^2}{\sum^{+N}_{l=-N}\left( S^{\rm in}_l - \langle S^{\rm in}_l\rangle \right)^2}\times 100 \%,
\end{equation}
where $S^{\rm in}_l$ and $S^{\rm ob}_l$ are the input and the observed OAM spectra, and $\langle S^{\rm in}_l \rangle$ is the average of $S^{\rm in}_l$.

Although in this section we have worked out the theory only for symmetric spectrum, in which case $S_l=S_{-l}$, our scheme works for non-symmetric spectrum as well (see Section 5 of the Supplementary Materials). However, for a non-symmetric spectrum, our scheme requires four intensity measurements instead of just two in the case of a symmetric spectrum. So, in situation in which one has the prior information that the spectrum is symmetric, one can measure the spectrum using only two intensity measurements. Otherwise, using four intensity measurements, one can measure any spectra, symmetric or asymmetric.

\begin{figure*}[!t]
\centering
\includegraphics[scale=0.84]{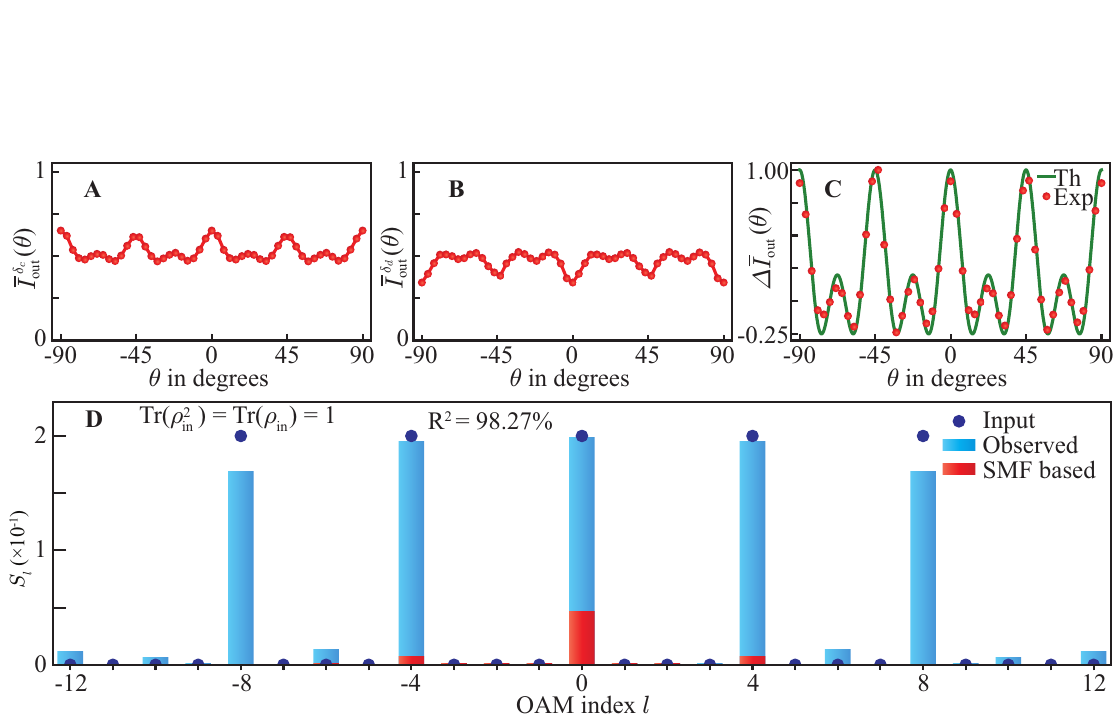}
\caption{{\bf Experimentally measured OAM probability distribution of discrete pure state.} {\bf(A)} and {\bf(B)} are the plots of the photon detection probability $\bar{I}_{\rm out}^\delta(\theta)$ as a function of the rotation angle $\theta$ at $\delta= \delta_{c} \approx 0$ and $\delta= \delta_{d} \approx  \pi$. {\bf(C)} Experimentally measured polarization-corrected difference probability $\Delta I _{\rm out}\left( \theta \right)$ as a function of $\theta$. The red dots are the experiments and the solid green line is the theory. {\bf(D)} Reconstructed OAM spectrum. Blue bars are the measured spectrum while the blue dots are the input spectrum; the red bars represent the performance of an SMF-based detector.}
\label{fig:pure_state_comb_result}
\end{figure*}

\begin{figure*}[!t]
\centering
\includegraphics[scale=0.84]{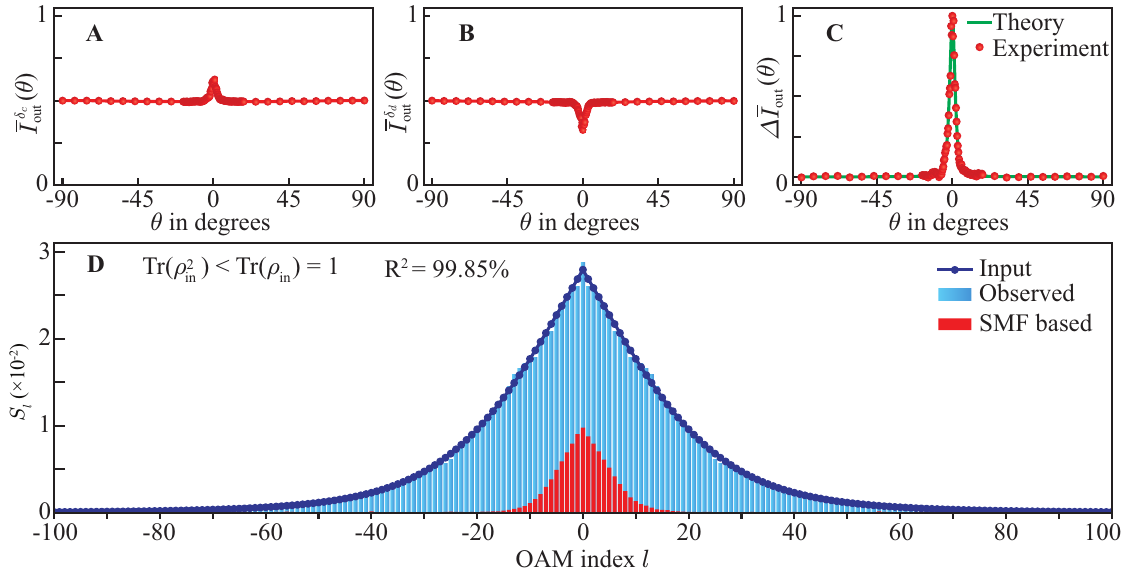}
\caption{{\bf OAM probability distribution of a single-photon diagonal mixed state produced by SPDC.} {\bf(A)} and {\bf(B)} are the plots of the photon detection probability $\bar{I}_{\rm out}^\delta(\theta)$ as a function of the rotation angle $\theta$ at $\delta= \delta_{c} \approx 0$ and $\delta= \delta_{d} \approx  \pi$. {\bf(C)} Experimentally measured polarization-corrected difference probability $\Delta I _{\rm out}\left( \theta \right)$ as a function of $\theta$. The red dots are the experiments and the solid green line is the theory. {\bf(D)} Reconstructed OAM spectrum. Blue bars are the measured spectrum while the blue dots are the input spectrum; the red bars represent the performance of an SMF-based detector. }
\label{fig:spdc_diagonal_mixed_state_result}
\end{figure*}

\noindent {\bf Experimental setup} \\
We employ two separate methods for generating a variety of single-photon quantum states in the OAM basis. Figure~{\ref{fig_exp_setup}}A is the schematic of the experimental setup for generating states using an attenuated He-Ne laser and an SLM while Fig.~{\ref{fig_exp_setup}}B is that using the nonlinear optical process of spontaneous parametric down-conversion (SPDC). The generated states are detected using the proposed OAM detector, the schematic setup of which is shown in Fig.~{\ref{fig_exp_setup}}C. The setup involves an EMCCD camera for ensuring that the detection is at the single-photon level. We report experimental measurements for various different types of single-photon OAM states in Figs~\ref{fig:gaussian_pure_state_result}, \ref{fig:gaussian_gen_mixed_state_result}, \ref{fig:pure_state_comb_result}, and \ref{fig:spdc_diagonal_mixed_state_result}. In each of these figures, the sub-figures A and B show the plots of  $\bar{I}_{\rm out}^\delta(\theta)$ as a function of $\theta$ at $\delta= \delta_{c} \approx0$ and $\delta_{d}\approx \pi$. The sub-figure C presents the measured polarization-corrected difference probability $ \Delta I _{\rm out}\left( \theta \right)$; the red dots are the experiments and the solid green line is the theory. Using  equation (\ref{eqn:normalization}), we calculate $S_l$ from the measured $\Delta I _{\rm out}\left( \theta \right)$ and plot them in sub-figure D, where the blue bars are for the observed and the blue dots are for the input states. The red bars represent the performance of the SMF-based method \cite{mair2001nature, heckenberg1992optlett} in the same situation with an SLM that has 50$\%$ diffraction efficiency in the first order. The measurement accuracy $R^2$ is shown inside each sub-figure D. (see Materials and Methods section for details on experimental techniques, procedures, analysis, etc.) \\

\noindent {\bf Generating quantum states using an attenuated He-Ne laser and an SLM}\\
For generating OAM states using the experimental setup of Fig.~{\ref{fig_exp_setup}}A, we use a Newport He-Ne laser of wavelength $633$ nm followed by an SLM. Using the method by Arrizon {\it et al.} \cite{arrizon2007josaa}, we generate various input states $\rho_{\rm in}$. First, we generate a pure quantum state with its OAM spectrum in the form of a Gaussian function of standard deviation equal to $8$. Figure~\ref{fig:gaussian_pure_state_result} presents the experimental results. We obtain the measurement accuracy $R^2$ of $98.96\%$. Next, we generate a general mixed state by incoherently adding two pure states in $0.54$ and $0.46$ proportion \cite{mandel_and_wolf1995}; the pure states have Gaussian OAM spectra with standard deviations equal to $10$ and $6$. For details on how a general mixed state is produced using an SLM, see Ref.~\cite{bhattacharjee2019jopt}. Figure \ref{fig:gaussian_gen_mixed_state_result} shows the experimental results, with $R^2 = 98.58\%$. Finally, we demonstrate that our OAM detector is not only capable of detecting a continuous set of OAM modes but also a discrete set. For this purpose, we generate a state with five OAM modes ($l=0,\pm4, \pm8$) with equal weightage and present the experimental results in Fig.~\ref{fig:pure_state_comb_result}, with $R^2=98.27\%$.  
\\

\noindent {\bf Generating quantum states using SPDC} \\
Although one can synthesize an arbitrary single-photon quantum states using an attenuated He-Ne laser and an SLM, the implementation always introduces some generation errors. Such errors can be avoided if a quantum state is generated through some natural process. SPDC, a second-order nonlinear effect, is one such process which naturally generates single-photon quantum states in the form of diagonal mixed state. So, for generating diagonal mixed states, we use SPDC, in which a pump photon gets down-converted into two separates photons called the signal and idler \cite{karan2020jopt}, and for a Gaussian pump field, the states of both the signal and idler photons are diagonal mixed states in the OAM basis \cite{jha2011pra, pires2010prl}. The spectrum $S_l$ of each of the two photons depends on several experimental parameters  such as  the pump beam waist $w_0$, crystal thickness $L$, phase-matching angle $\theta_p$ etc. \cite{karan2020jopt, kulkarni2018pra}. Using SPDC in a $\beta$-barium borate (BBO) crystal with collinear phase-matching condition, we experimentally generate single-photon diagonal mixed states using the setup depicted in Fig.~{\ref{fig_exp_setup}}B, with $\theta_p= 28.668^{\circ}$, $L= 15$ mm and  $w_0=388$ $\mu$m.  Figure~\ref{fig:spdc_diagonal_mixed_state_result} presents the experimental results, with $R^2=99.85\%$. We find that for the same state, the detection efficiency of the SMF-based measurement scheme is relatively much less, is mode dependent, and becomes negligible for OAM modes with $|l|> 20$.\\

\begin{figure}[t!]
\centering
\includegraphics[scale=1]{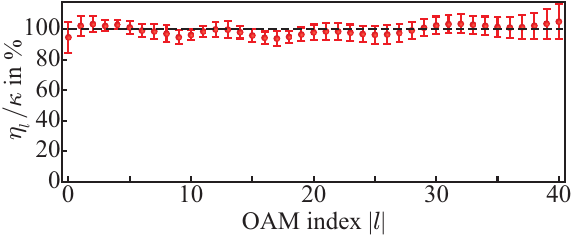}
\caption{{\bf Estimated detection efficiency as a function of $|l|$.}} 
\label{fig:each_mode_efficiency}
\end{figure}

\noindent {\bf Note on the generation error}\\
We note that the measurement accuracy of the results reported in Figs.~\ref{fig:gaussian_pure_state_result}D, \ref{fig:gaussian_gen_mixed_state_result}D, and \ref{fig:pure_state_comb_result}D is smaller than that of Fig.~\ref{fig:spdc_diagonal_mixed_state_result}D. We attribute this to the error in the generations process adopted to produce the states. For the results reported in Figs.~\ref{fig:gaussian_pure_state_result}D, \ref{fig:gaussian_gen_mixed_state_result}D, and \ref{fig:pure_state_comb_result}D, the state was generated using an SLM. The input spectrum plotted as blue dots in Figs.~\ref{fig:gaussian_pure_state_result}D- \ref{fig:pure_state_comb_result}D, is what we ideally desired to generate but it is not what was generated by the SLM due to all the associated generation errors. Nonetheless, we plotted the blue dots as input spectrum. Therefore, the inaccuracies in Figs.~\ref{fig:gaussian_pure_state_result}D- \ref{fig:pure_state_comb_result}D are primarily because of the error in the generation process and thus in the plotted input spectrum but not because of the error in the observed spectrum. However, we note that, if the state generation has no error, the match between the input and observed spectra is much better. This can be observed through the results presented in Fig.~\ref{fig:spdc_diagonal_mixed_state_result}D. In this case, since the input state was naturally produced using the nonlinear optical process of SPDC, there was no generation error and it could be very accurately predicted by the theory. As a consequence, the measurement accuracy for this data set is higher. We further note that because of the issues related to generation accuracy, we only use data presented in Fig.~\ref{fig:spdc_diagonal_mixed_state_result}D for estimating the detection efficiency presented in Fig.~\ref{fig:each_mode_efficiency}.\\

\noindent {\bf Analyzing the uniformity of detection efficiency}\\
One of the objectives of this work is to develop a detector that has uniform detection efficiency over a broad range of OAM modes as opposed to the SMF-based detectors \cite{mair2001nature, heckenberg1992optlett}. We take the mode detection efficiency $\eta_l$ of our scheme to be:
\begin{align}
\eta_l =\kappa \ \frac{S^{\rm ob}_l}{S^{\rm in}_l} \times 100\%,
\end{align}
where $\kappa$ is the overall quantum efficiency of the detection system. $S^{\rm in}_l$ and  $S^{\rm ob}_l$ are the input and the observed OAM spectra. Since the results reported in Fig.~\ref{fig:spdc_diagonal_mixed_state_result}D have the lowest state generation error, we use these for estimating $\eta_l$ and plot it in Fig.~\ref{fig:each_mode_efficiency}. The factor $\eta_l/\kappa$ is uniform over $|l|$ demonstrating a broadband mode-independent detection efficiency. The increase in the variance of $\eta_l/\kappa$ at large $|l|$ is attributed to decreasing signal-to-noise ratio of the data set at large $|l|$ which eventually becomes unfit for estimating $\eta_l$
beyond $|l|=40$.

\section*{\label{sec:Discussion} Discussion}

In conclusion, we have proposed and demonstrated a broadband, uniform efficiency OAM-mode detector that measures the true OAM-mode spectrum, is noise-insensitive, and is capable of measuring the OAM spectrum of an arbitrary quantum state without any prior information both at high-light and single-photon levels. Such an OAM-mode detector can have important consequences for OAM-based photonic quantum technologies, such as the high dimensional cryptography protocols \cite{mirhosseini2015njp} and optical communications \cite{bozinovic2013science}, quantum computing \cite{babazadeh2017prl}, quantum metrology \cite{lavery2013science, cvijetic2015scirep}, and quantum imaging \cite{chen2014lsa,torner2005oe}. We believe that the OAM-mode detector reported here will not only render the existing OAM-based applications more viable but also enable novel OAM-based quantum technologies.

In order to emphasize the noise-insensitive aspect of our technique, we note that in sub-figures A and B of Figs.~\ref{fig:gaussian_pure_state_result}-\ref{fig:spdc_diagonal_mixed_state_result}, the plots of $\bar{I}_{\rm out}^{\delta_c}(\theta)$ and $\bar{I}_{\rm out}^{\delta_d}(\theta)$ contain large background contributions,  resulting in poor contrast, the main sources of which are the EMCCD dark counts and the ambient light. This is because in our experiment, the transverse size of the field falling onto the EMCCD is much smaller than the EMCCD sensor size. As a result, only a small fraction of the EMCCD pixels is involved in detecting the field and contributing to the signal; the remaining pixels only contribute to noise. Nonetheless, as shown in sub-figures C, the noise contributions  $ I_n^{\delta_c}(\theta)$ and $I_n^{\delta_d}(\theta)$ cancel each other in the difference probability $\Delta\bar{I}_{\rm out}(\theta)$ which becomes almost free of such noise. As a result, our technique becomes extremely insensitive to noise sources and thus achieves very high detection fidelity even in the presence of substantial noise. This is in contrast to the most widely used method of OAM-mode detection based on an SLM and an SMF  \cite{mair2001nature, heckenberg1992optlett} that cannot filter out such noise sources and thus get adversely affected by them.

In addition to being  noise-insensitive, our technique does not involve loss of the input state and can be very fast. A careful alignment of our home-built IR brings the angular deviations to about 30 $\mu$-radians, and we observed that a realignment of IR was required only if $\theta$ was changed by more than 30$^{\circ}$. Therefore, if there is prior information that the angular width of $\bar{I}^{\delta}_{\rm out}\left(\theta\right)$ is less than 30$^{\circ}$ then no realignment is required. However, if we have no prior information or if $\bar{I}^{\delta}_{\rm out}\left(\theta\right)$ has larger angular widths, realignments are required. For the spectra reported in Figs~\ref{fig:gaussian_pure_state_result}, \ref{fig:gaussian_gen_mixed_state_result}, and \ref{fig:spdc_diagonal_mixed_state_result}, no realignment was required and the entire data taking was finished within a few minutes. However, for the spectra in   Fig.~\ref{fig:pure_state_comb_result}, realignments were done, increasing the data taking time proportionately.

We note that our new approach for broadband OAM-mode detection ensures unique features such as (i) uniform detection efficiency, (ii) true spectrum measurement without post-selecting on a particular radial mode, (iii) no need for prior information, and (iv) insensitivity to phase-uncorrelated background. However, our present implementation technique that has been demonstrated for the diagonal elements of an arbitrary density matrix in the OAM basis, cannot be employed for measuring the off-diagonal elements which contains information about whether two given OAM modes are phase-coherent or phase-incoherent with respect to each other. Nonetheless, if such an approach could be extended to off-diagonal elements, then for an arbitrary density matrix even the off-diagonal elements could be measured with high-fidelity and the measurement scheme by default would have all the unique features of this approach.

The prime reason for stringent alignment and stability requirements in our technique is the unavailability of IRs with near-zero angular deviations and faster rotation speeds. If such image rotators become commercially available, the alignment and stability requirements as well as the measurement time will reduce proportionately, making our technique suitable even for applications that requires very fast measurements. Until such IRs become commercially available, the existing techniques such as those based on an SLM and an SMF may be preferred in situations in which the dimensionality of the state is very low and the true state measurement with high fidelity is not crucial. However, for any high-dimensional applications requiring accurate, noise-robust, and high-fidelity measurement of OAM spectrum, our technique would be most suited. Also, although our technique is not based on having prior knowledge of the input state, leveraging such knowledge, as is done in compressive sensing based methods \cite{tonolini2014scirep}, can further reduce the alignment requirements and significantly increase the measurement speeds.

\section*{\label{sec:materials and methods} Materials and Methods}

\begin{figure}[h!]
\centering
\includegraphics[scale=1]{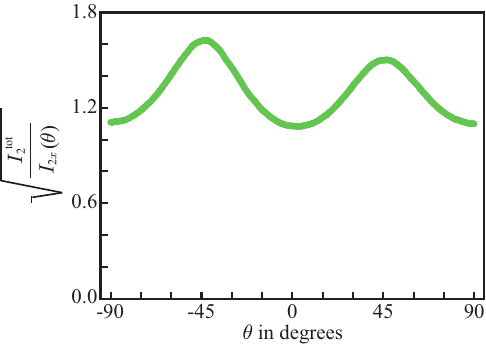}
\caption{{\bf Measured $\frac{1}{\cos[\psi(\theta)]}=\sqrt{\frac{I^{\rm tot}_2}{I_{2x}(\theta)}}$ as a function of $\theta$.} }
\label{fig:polarization_calibration}
\end{figure}

\noindent {\bf Polarization calibration of the interferometer}\\
In order to bypass the polarization issues caused by the rotation of the IR, the interferometer needs to be calibrated according to equation (\ref{deltaI_pol_corrected}) by measuring both  $I_2^{\rm tot}$ and $I_{2x}(\theta)$ and thus $\cos[\psi(\theta)]$. The probabilities $I_2^{\rm tot}$ and $I_{2x}(\theta)$ can be measured by blocking the upper arm of the interferometer. For measuring $I_{2x}(\theta)$, we place a polarizer along $\hat{\bm x}$, right before the EMCCD, and measure the probability as a function of $\theta$. We note that $I_2^{\rm tot}$ is the total probability through the IR without any polarizer and thus that it is independent of $\theta$. Figure~\ref{fig:polarization_calibration} shows the plot of $\dfrac{1}{\cos[\psi(\theta)]}=\sqrt{\frac{I^{\rm tot}_2}{I_{2x}(\theta)}}$ and is used for calculating $\Delta I_{\rm out} \left(\theta\right)$ using equation (\ref{deltaI_pol_corrected}) for the results reported in Figs~\ref{fig:gaussian_pure_state_result}, \ref{fig:gaussian_gen_mixed_state_result}, \ref{fig:pure_state_comb_result}, and \ref{fig:spdc_diagonal_mixed_state_result}.

\noindent{\bf Details of the experimental setup in Fig.~{\ref{fig_exp_setup}}A}\\
In this experimental setup, we generate  high-dimensional OAM states using a He-Ne laser of wavelength $633$ nm and a beam waist of $0.4$ mm. We make the Gaussian beam go through a spatial filter and then magnify it using a $4f$ configuration with lenses of focal lengths $13.86$ mm and $200$ mm. The resulting beam is projected onto a Holoeye Pluto-2-NIR-011 SLM. We use the method by Arrizon et al.\cite{arrizon2007josaa} to imprint a suitable hologram on the SLM for producing a desired coherent superposition of $LG ^{|l|}_{p=0}$ modes. The generated state is obtained at the first diffraction order of the SLM. To measure the probability, we use an Andor iXon Ultra897 EMCCD camera with a $512 \times 512$ pixel grid and a pixel pitch of $16 \times 16$ $\mu{\rm m}^2$. EMCCD gain is set at $1$ with sufficient neutral density filter in order to avoid saturation.

\

\noindent{\bf Details of the experimental setup in Fig.~{\ref{fig_exp_setup}}B}\\ In this setup, we generate, single-photon states by producing SPDC photons in a type-I BBO crystal of thickness $15$ mm, pumped by a UV $100$ mW TOPTICA TopMode continuous wave laser of wavelength $ 405$ nm with beam waist $388$ $\mu$m. A dichroic filter is used for blocking the UV laser source while allowing the down-converted photons at $810$ nm to pass through. Subsequently, the down-converted photons are filtered through a $10$ nm bandwidth interference filter having a central wavelength of $810$ nm. The crystal plane is imaged with a magnification of 6 using the $4f$ configuration with lenses of focal lengths $50$ mm and $300$ mm,. The Fourier transform of this magnified image is obtained at the EMCCD plane using a lens of focal length $500$ mm. The EMCCD is used with a gain of 300 and an acquisition time of 0.4 seconds.

We note that since we are using collinear down-conversion, the individual signal and idler photons have equal probability of arriving at the EMCCD. Therefore, the camera records the sum of photon count rates due to signal and idler fields. However, since the individual signal and idler fields have the same OAM spectrum, the difference probability $\Delta I_{\rm out}(\theta)$ due to the signal and idler photons are also the same. Here, we have assumed that  the probability of the simultaneous arrival of the signal and idler photons at the same EMCCD pixel is negligibly small. This assumption is justified given that the EMCCD has 512 × 512 pixels, and we work at the level of only a few photons per EMCCD frame.

\

\noindent {\bf Aligning the interferometer in Fig.~{\ref{fig_exp_setup}}C}\\
In our experiment, first we align the IR manually to reduce the angular deviations of the field passing through it to be  less than $1$ milli-radian. We then insert two mirrors $M_1$ and $M_2$ with actuator-based control after the IR and use a feedback mechanism for automated alignment. To track the center of the Gaussian beam from the He-Ne laser  at two locations, we use two cameras placed at the output ports of the beam splitter. One camera is a charge-coupled device (CCD) camera of IDS imaging with a pixel grid of $2592 \times 2048$, while the other is the EMCCD camera. Next, we manually align the interferometer for zero fringe condition with a Gaussian beam and block the upper interferometric arm containing Q-H-Q to determine the coordinates of the center of the beam falling on each camera. These coordinates serve as the target coordinates for alignment at each subsequent $\theta$, if re-alignment is required. We then center the beam by electronically adjusting the mirrors $M_1$ and $M_2$ through an iterative algorithm. For this, we make sure that the distance between the EMCCD and BS$_2$ is far greater than that between the CCD and BS$_2$. Then, by adjusting $M_1$, we bring the beam towards the CCD target center, and by adjusting $M_2$, we bring the beam towards the EMCCD target center. This way, we are able to align the fields to within $30.5$ $\mu$-radian to the EMCCD. To set the interferometer for $\delta = \delta_c$ and $\delta=\delta_d$ conditions, we remove the blocker from the upper arm and record the output intensity $I$ at the EMCCD camera for a Gaussian input beam. We take five such intensity measurements at five different orientations $\beta$ of the half-wave plate, separated by $36^{\circ}$. We fit these  measurements with the function $I = a + b\cos 2\left(\beta - c\right)$ and obtain the orientation at which $I$ is maximum and thus obtain $\delta \approx 0$. From this orientation, we rotate the half-wave plate by $90^{\circ}$ to obtain the $\delta \approx \pi$ condition. We then measure the probabilities $\bar{I}^{\delta_{c}}_{\rm out}\left(\theta\right)$ and $\bar{I}^{\delta_{d}}_{\rm out}\left(\theta\right)$. To reduce the effects due to thermal and air-flow fluctuations, we cover the interferometer with a box and wait for it to stabilize.

\

\noindent {\bf Rotation resolution and OAM spectrum reconstruction}\\
Our technique requires $S_l$ to be reconstructed from the  measured $\Delta I_{\rm out} \left(\theta\right)$ through Fourier transformation. According to the whittaker-Shannon sampling theorem \cite{goodman2005}, the sampling rate of a signal should be at least twice its highest frequency to completely characterize the signal. So, for our scheme, if the highest OAM index of the incident state is $l_{\rm max}$, the step size for $\theta$ should be smaller than $180^{\circ}/(2 l_{\rm max} +1)$. Therefore, in order to reconstruct an OAM spectrum having $l_{\rm max}$ number of modes, one needs to measure $\Delta I_{\rm out} \left(\theta\right)$ at a minimum of $2 l_{\rm max}$ number of $\theta$ points. Nonetheless, since we do not use any prior information about the input state including its dimensionality, we first measure $\Delta I_{\rm out} \left(\theta\right)$ at fewer numbers of $\theta$ values to get some idea about the functional form of $\Delta I_{\rm out} \left(\theta\right)$, and then repeat the  measurements of $\Delta I_{\rm out} \left(\theta\right)$ at an appropriate number of $\theta$ values as dictated by the sampling theorem. For the state reported in Figs.~\ref{fig:gaussian_pure_state_result} and ~\ref{fig:pure_state_comb_result}, we measure $\Delta I_{\rm out} \left(\theta\right)$ at $\theta$ values separated by $3.6^{\circ}$. For the state reported in Figs.~\ref{fig:gaussian_gen_mixed_state_result} and \ref{fig:spdc_diagonal_mixed_state_result}, the variation of $\Delta I_{\rm out} \left(\theta\right)$ is maximum in the range $\theta=-18^{\circ}$ to $\theta=18^{\circ}$. So, for this range, we take measurements at $\theta$ values separated by $0.36^{\circ}$, while for the remaining range, the separation is $7.2^{\circ}$.

\

\noindent {\bf Effects due to the angular deviation of the image rotator}\\
In our technique, it is very important that the output intensity $\bar{I}_{\rm out}^\delta(\theta)$ as a function of rotation angle $\theta$ is measured correctly. And for this, it is necessary that the two fields interfering at the second beam splitter (BS$_2$) of Fig.~1C have perfect overlap; any deviation from the perfect overlap results in error in the measurement of $\bar{I}_{\rm out}^\delta(\theta)$ and thus in the estimated spectrum. However, due to manufacturing errors, IRs invariably introduce angular deviations in the transmitted field resulting in decreased overlap as a function of $\theta$. In Section 6 of the Supplementary Material, we present a detailed analysis of the effects due to angular deviation. In that section, through illustrative figures and numerical analysis, we show how angular deviation of an image rotator affects the measurement of $\bar{I}_{\rm out}^\delta(\theta)$. We find that  with a commercial Dove prism having angular deviation greater than 10000 $\mu$-radians, the average fractional overlap of the two fields is less than 0.1, which implies a very inaccurate measurement of $\bar{I}_{\rm out}^\delta(\theta)$. On the other hand, for the home-built image rotator reported in this work, having angular deviations less that 30 $\mu$-radians, the average fractional overlap is close to 1, resulting in a very accurate measurement of $\bar{I}_{\rm out}^\delta(\theta)$.

\

\begin{figure}[b!]
\centering
\includegraphics[scale=0.45]{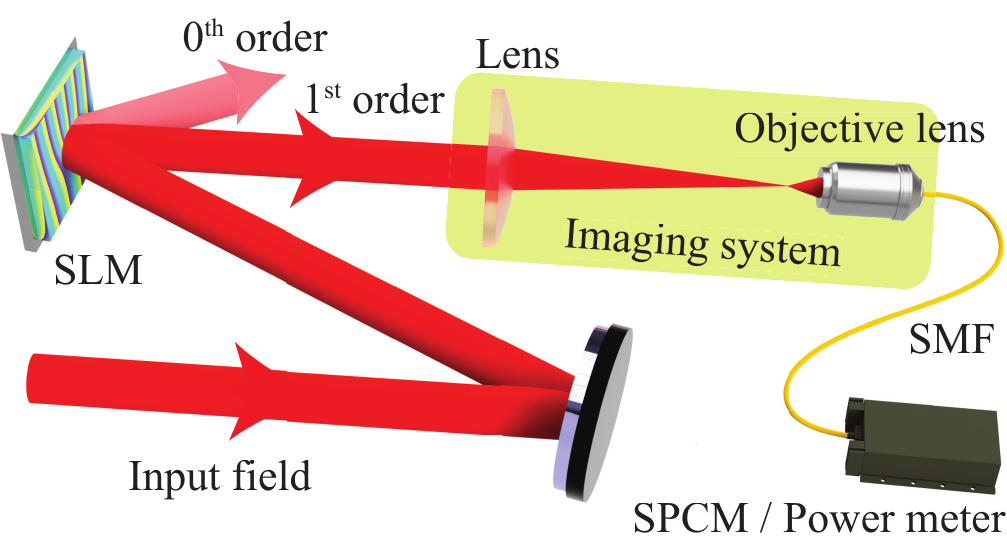}
\caption{{\bf Experimental setup for OAM mode detection using an SLM and an SMF.} SPCM: single photon counting module.}
\label{fig:fiber_based_exp_setup}
\end{figure}

\noindent {\bf Details of the experimental setup for single-mode fiber-based OAM measurement}\\
Figure~\ref{fig:fiber_based_exp_setup} shows the experimental setup for an SMF-based measurement \cite{mair2001nature, heckenberg1992optlett}. For the input spectra produced using the setup of Fig.~{\ref{fig_exp_setup}}A and Fig.~{\ref{fig_exp_setup}}B, we couple the first-order diffraction output from the SLM to the SMF by imaging the SLM plane onto the SMF using a $500$ mm focal length lens and a Newport $60\times$ objective  with effective focal length $2.9$ mm in a $4f$ configuration. In this configuration, we obtain $\sigma/w_0=0.6$, where $\sigma=M\sigma_{\rm smf}$, $w_0$ is the beam-waist of the OAM modes, $\sigma_{\rm smf}$ is the core radius of the SMF, and $M$ is the magnification of the the imaging system. The SMF is then coupled to a single-photon avalanche diode detector to measure the OAM spectrum. For the state generated using the setup of Fig.~{\ref{fig_exp_setup}}B, we image the BBO crystal of Fig.~{\ref{fig_exp_setup}}B onto the SLM.

\bibliography{ref_oam_detector_manuscript}

\bibliographystyle{sciencemag}

\section*{Acknowledgments}

We thank Sanjana Wanare and Swadha Pandey for their initial contributions in implementing the image rotator. 
\\

\noindent{\bf Funding:}
We acknowledge financial support from the Science and Engineering Research Board through grants STR/2021/000035 and CRG/2022/003070 and from the Department of Science $\&$ Technology, Government of India through Grant DST/ICPS/QuST/Theme-1/2019). SK thanks the University Grant Commission (UGC), Government of India for financial support.
\\

\noindent{\bf Author Contributions:}
AKJ and SK proposed and developed the idea. SK performed the experiments with help from AKJ. MPVE provided the high-quality home-built IR. AKJ and SK wrote the manuscript with inputs from MPVE. AKJ supervised the overall work.
\\

\noindent{\bf Competing interests: }
The authors declare that they have no competing interests.
\\

\noindent{\bf Data and materials availability: }
All data needed to evaluate the conclusions in the paper are present in the paper and/or the Supplementary Materials.

%%%%%%%%%%%%%%%% SUPPLEMENT LIST %%%%%%%%%%%%%%%

% List the contents of your Supplementary Materials, including the numbers of any
% supplementary figures, tables, external data files etc. and any references that are
% cited only in the supplement. In this example, refs. 7-8 are cited only in the supplement.
% Fill out your numbers accordingly and delete any lines that aren't applicable.
\subsection*{Supplementary materials}
Supplementary Text\\
Figs. S1 to S5\\
%%%%%%%%%%%%%%%% END OF MAIN TEXT %%%%%%%%%%%%%%%

\newpage

%%%%%%%%%%%%%%%% START OF SUPPLEMENT %%%%%%%%%%%%%%%

% Figures, tables, equations and pages in the supplement are numbered S1, S2 etc.
\renewcommand{\thefigure}{S\arabic{figure}}
\renewcommand{\thetable}{S\arabic{table}}
\renewcommand{\theequation}{S\arabic{equation}}
\renewcommand{\thepage}{S\arabic{page}}
\setcounter{figure}{0}
\setcounter{table}{0}
\setcounter{equation}{0}
\setcounter{page}{1} % not 0 as \newpage already started a supplementary page
% References continue the numbering from the main text.

%%%%%%%%%%%%%%%% SUPPLEMENT TITLE PAGE %%%%%%%%%%%%%%%

\begin{center}
\section*{Supplementary Materials for\\ \scititle}

% Author list for the supplement
% Indicate the corresponding authors, but do NOT include institutions here
% It would be nice if the template auto-generated this, but doing so is complicated...
Suman Karan$^{\ast}$,
Martin P. Van Exter,
and Anand K. Jha$^{\ast}$\\ % we're not in a \author{} environment this time, so use \\ for a new line
\small$^\ast$Corresponding author. Email: karans@iitk.ac.in, akjha@iitk.ac.in.
\end{center}

\subsubsection*{This PDF file includes:}
Supplementary Text\\
Figs. S1 to S5\\

\newpage

%%%%%%%%%%%%%%%% SUPPLEMENTARY TEXT %%%%%%%%%%%%%%%

\subsection*{Supplementary Text}

\section{Mode dependent detection efficiency in sigle-mode fiber-based measurement}
\begin{figure}[b!]
\centering
\includegraphics[scale=0.94]{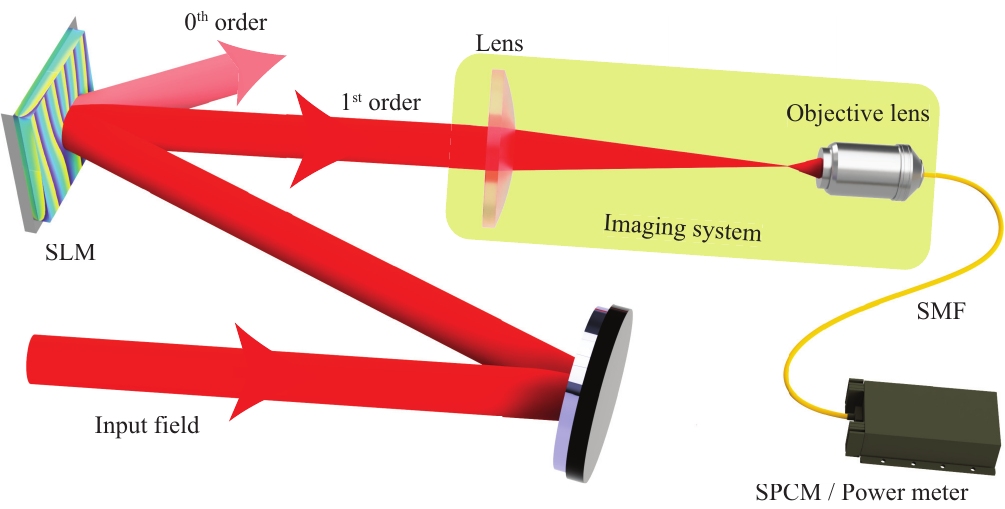}
\caption{{\bf Experimental setup for OAM mode detection using a spatial light modulator (SLM) and a single-mode fiber (SMF).} SPCM: single photon counting module. }
\label{fig:supp_fiber_based_exp_setup}
\end{figure}
Figure~\ref{fig:supp_fiber_based_exp_setup} shows the typical experimental setup for the single-mode fiber-based (SMF) OAM spectrum measurement \cite{mair2001nature, heckenberg1992optlett} . The electric field of a Laguerre-Gaussian (LG) mode  $LG^{l}_{p} (\rho, \phi) $ at the beam waist plane can be written as \cite{allen1992pra} %\cite{mair2001nature, heckenberg1992optlett}   %\cite{allen1992pra}
\begin{equation}
LG^{l}_{p} (\rho, \phi) =\sqrt{\dfrac{2~ p!}{\pi {w_0}^2 \left(p+|l|\right)! }}\left(\dfrac{\rho \sqrt{2}}{w_0}\right)^{|l|} {\rm exp}\left[-\dfrac{\rho^2}{w^2_0}\right] L^{|l|}_p\left(\dfrac{2 \rho^2}{w^2_0}\right) e^{-i l \phi},
\end{equation}
where $l$ is an integer ranging from $-\infty$ to $\infty$ and is referred to as the OAM-mode index while $p$ is called the radial-mode index, and it ranges from $0$ to $\infty$. $L^{l}_p[\dots]$ is called the  associated Laguerre polynomial given by $L^{l}_p[x]= \sum^{p}_{m=0}(-1)^{m}\dfrac{(p+l)!}{(p-m)! (l+m)!}\dfrac{x^{m}}{m!}$ and $w_0$ is the beam waist of the field. The field amplitude $LG^{l}_{p} (\rho, \phi)$ can be written as a product of two separate parts, one containing the helical phase  and the other containing the radial dependence: $LG^{l}_{p} (\rho, \phi)=LG^{|l|}_{p}(\rho)e^{-i l\phi} $. For detecting this mode using an SMF, we need to first get rid of the helical phase $e^{-i l\phi}$. For this, an additional complementary phase  $e^{il \phi}$ is introduced by using a spatial light modulator (SLM). Next, as shown in Fig.~\ref{fig:supp_fiber_based_exp_setup}, the first-order diffracted field from the SLM is coupled into the SMF of core radius $\sigma_{\rm smf}$ thorough a $4f$ lens configuration and collected either using a powermeter for high-light level measurement or using a single-photon counting module (SPCM) for single-photon counting. The SMF tip is imaged onto the SLM with a magnification factor of $M$. Therefore, the fundamental mode of the single-mode fiber (SMF) at the SLM plane can be written as 
\begin{equation}\label{eqn:smf mode}
G(\rho)= \sqrt{\dfrac{2}{\pi \sigma^2}} e^{- \dfrac{\rho^2}{\sigma^2}},
\end{equation}
where $\sigma= M \sigma_{\rm smf}$ is the mode radius at the SLM plane. Thus, the coupling-coefficient to the SMF for the input mode $LG^{l}_{p} (\rho, \phi) $ can be obtained by the following overlap integral 
\begin{equation}\label{eqn: coeff_efficiency}
C^p_l = \int^{\infty}_{\rho=0}\int^{2\pi}_{\phi=0} LG^{l}_p(\rho, \phi) \left[e^{-i l \phi}\right]^* \sqrt{\dfrac{2}{\pi \sigma^2}} e^{- \rho^2/\sigma ^2} \rho d\rho d \phi.
\end{equation}
\begin{figure}[t!]
\centering
\includegraphics[scale=0.9]{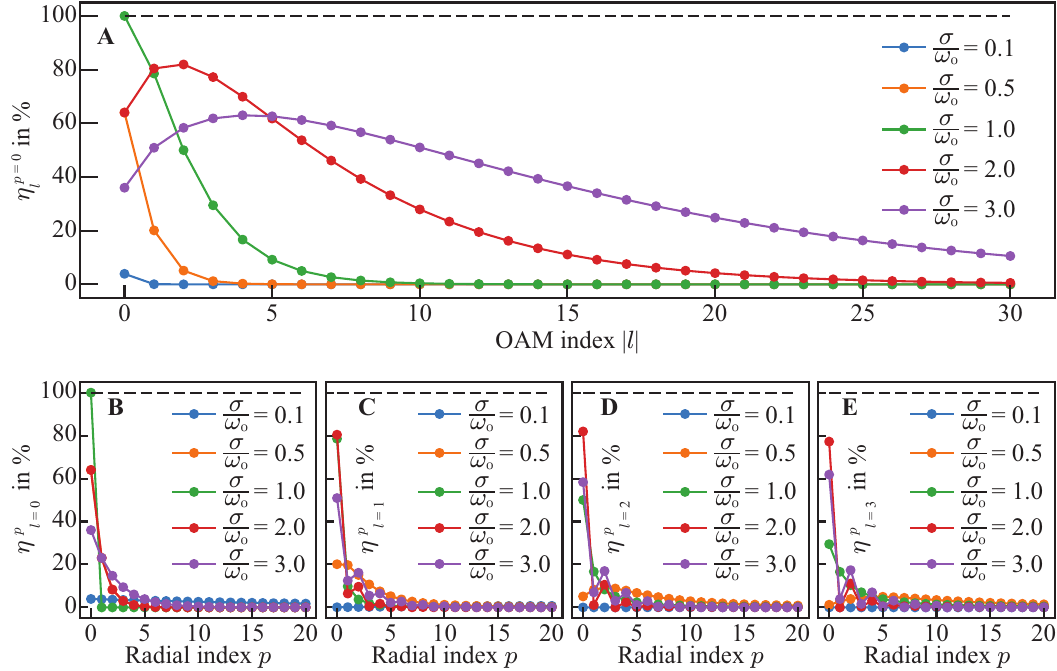}
\caption{{\bf Coupling efficiency as a function of OAM mode index $l$  and radial mmode index $p$ for five different values of the ratio  $\sigma/w_0$.} {\bf(A)} Plot of the coupling efficiency $\eta_{l}^0$ as a function of $l$. {\bf (B-E)} Plots of $\eta^{p}_l$ as a function of  $p$ for $l=0$, $l=1$, $l=2$, and $l=3$, respectively.  }
\label{fig:effi_oam_mode_plt}
\end{figure}
The magnitude $|C^p_l|^2$ is taken as the coupling efficiency. In practical experimental situations, the  first order diffraction efficiency of the SLM and the quantum efficiency of the detectors (SPCM/powermeter) are never perfect, and as a result, the overall quantum efficiency of the detection system $\kappa$ is always less than $1$. The detection efficiency of the system for an LG mode can be defined as $\eta^{p}_l= \kappa |C^p_l|^2$, and after evaluating the overlap integral above, one can obtain the following expression for the detection efficiency $\eta^{p}_l$:
\begin{equation}\label{eqn:efficiency}
\eta^{p}_l= \kappa \dfrac{4\pi (p+|l|)!}{2^{|l|}p!}\left(\dfrac{w_0}{\sigma} \right)^2\left| \left[ \dfrac{(1+ \frac{w_0^2}{\sigma^2})^{-1- |l|/2}}{\Gamma{\left(\frac{|l| + 1}{2}\right)}}\right] 2F1\left[ -p, 1 + \frac{|l|}{2}, 1+ |l|, \frac{2}{1+ \frac{w_0^2}{\sigma^2}} \right] \right|^2.
\end{equation}
Here $2F1[\dots]$ represents the Gaussian hypergeometric function. In Fig.~\ref{fig:effi_oam_mode_plt}A we plot $\eta^{0}_{l}/\kappa$ as a function of the OAM mode index $|l|$, for various $\sigma/w_0$ values. We find that the detection efficiency of the SMF-based detector is highly dependent on the OAM mode index $|l|$ and the ratio  $\sigma/w_0$. The detection efficiency decreases very sharply with $|l|$, and as a result, the range of modes that can be detected with this detector is very small. Although this range can be increased by increasing $\sigma/w_0$, it leads to decrease in the overall efficiency of the detector. Figures.~\ref{fig:effi_oam_mode_plt} B-\ref{fig:effi_oam_mode_plt}E, show the plots of $\eta^{p}_{l}/\kappa$ as a function of the radial mode index $p$ for various $l$ and $\sigma/w_0$. We find that $\eta^{0}_{l}/\kappa$ is mostly maximum for $p=0$ mode. Thus, we see that the SMF-based OAM detectors can primarily detect $p=0$ mode, and even for this mode, these detectors have non-uniform detection efficiency and work for only a small range of OAM modes.

One might be inclined to think that by appropriately correcting for the fiber coupling efficiency, one can achieve uniform detection efficiency with an SLM-based OAM detector. However, we find that this may be possible to some extent, if one has the prior information that the spectrum-to-be-measured contains only $p=0$ modes. However, if the unknown state contains $p\neq0$ modes, this correction would be a very difficult task. As demonstrated in Figs.~\ref{fig:effi_oam_mode_plt} B-\ref{fig:effi_oam_mode_plt}E, different $p$ modes have different detection efficiencies. And unless one knows the proportion of these $p$  modes a priori, it would not be possible to correct for the coupling efficiencies. Furthermore, even when the OAM spectrum-to-be-measured contains only p=0 modes, it is not possible to implement the correction for more than a few modes. This is because, as discussed above and shown in Figs.~\ref{fig:effi_oam_mode_plt}, an SMF has an appreciable detection efficiency over only a small range of OAM modes. Beyond this range, the signal-to-noise becomes so low that the correction for the coupling efficiency would cause rapidly increasing error in the OAM spectrum estimation with the OAM index.

%%
%%
%\begin{figure}[hbtp]
%\centering
%\includegraphics[scale=1]{theory_fiber_measurement_gauss_pure.eps}
%\caption{Theoretical plot of the OAM spectrum for a pure state obtained using the SMF method with a $\dfrac{\sigma}{w_0}$ ratio of $0.6$. The red bar represents the measured OAM spectrum, and the blue dot corresponds to the input OAM spectrum.}
%\label{fig:th_pure_gauss_fiber_result}
%\end{figure}
%%
%%
%
%Figure~\ref{fig:th_pure_gauss_fiber_result} clearly illustrates this limitation. In this plot, we show the theoretically obtained OAM spectrum using the SMF-based method with a $\dfrac{\sigma}{w_0}$ ratio of $0.6$ for a pure state characterized by a Gaussian OAM spectrum with a standard deviation of $8$. This state corresponds to the one we used to demonstrate our method, as seen in Fig. 2 of the main text. Within the plot, the red bar represents the OAM spectrum obtained through the SMF-based method, while the blue dot signifies the true OAM spectrum.

\section{Mode transformation due to mirror reflection}\label{sec:mirror_operator}

\begin{figure}[b!]
\centering
\includegraphics[scale=1]{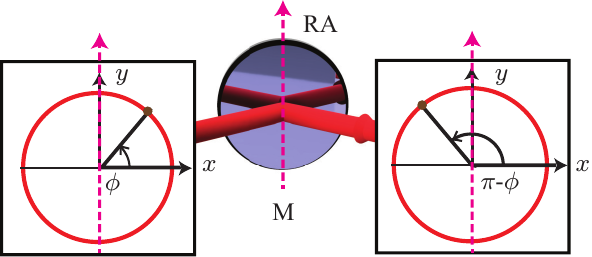}
\caption{ {\bf Mode transformation due to mirror reflection.}  An incoming field with transverse phase $e^{il\phi}$ transform into $e^{il(\pi-\phi)}$ through mirror reflection. RA is the reflection axis.}
\label{fig: mirror_transformation}
\end{figure}

The ket $|l_1, p_1\rangle$ represents the single-photon quantum state with mode indices ($l_1$, $p_1$) and $C^{p_1,p_2}_{l_1,l_2}$ is the density matrix element corresponding to indices ($(l_1, p_1)$) and ($(l_2, p_2)$). The projection of the state $|l_1, p_1\rangle$ on the transverse-polar basis state $|\rho,\phi\rangle$ is given by $\langle\rho,\phi|l_1,p_1\rangle = LG^{|l_1|}_{p_1}(\rho)e^{-il_1\phi}$. As shown in Figure~\ref{fig: mirror_transformation} , a mirror reflects an incoming field with azimuthal phase profile $e^{-il\phi}$ and transforms it into $e^{-il(\pi-\phi)}$. Thus, a mirror effectively transforms the state $|\rho, \phi\rangle $ into $|\rho, \pi- \phi\rangle $. This transformation by a mirror can be represented as $\hat{M} |\rho, \phi\rangle = |\rho, \pi- \phi\rangle$, where   $\hat{M}$ is the projection operator and can be written as:
\begin{align}\label{eqn:m_in_angle_basis}
\hat{M} &=\int^{\infty} _{0}\int^{2 \pi}_{0} \int^{\infty} _{0}\int^{2 \pi}_{0}  |\rho', \phi'\rangle \langle \rho', \phi'| \hat{M} |\rho, \phi\rangle \langle \rho, \phi| ~\rho\rho' d\rho d\rho' d \phi d \phi'. \notag \\
&=\int^{\infty} _{0}\int^{2 \pi}_{0} \int^{\infty} _{0}\int^{2 \pi}_{0}  |\rho', \phi'\rangle \langle \rho', \phi' |\rho, \pi-\phi\rangle \langle \rho, \phi| ~\rho\rho' d\rho d\rho' d \phi d \phi'. 
\end{align}
Using the inner product $ \langle \rho', \phi' |\rho, \pi-\phi\rangle=\frac{1}{\rho}\delta(\rho'-\rho)(\phi'-\pi+\phi)$, we write the projector as
\begin{align}
\hat{M}=\int^{\infty} _{\rho=0}\int^{2 \pi}_{\phi=0} |\rho, \pi- \phi\rangle \langle \rho, \phi| ~\rho d\rho ~d \phi.
\end{align}
Now, using the completeness condition $\sum_{l,p} |l,p\rangle \langle l,p| = \mathbb{I}$, we write $\hat{M}$  in the LG basis as 
\begin{equation} \label{eqn: intermidiate_M_phi_to_l}
\hat{M} = \sum_{l,p}\sum_{l', p'}\int^{\infty} _{\rho=0}\int^{2 \pi}_{\phi=0} |l',p'\rangle \langle l',p'| \rho, \pi- \phi\rangle \langle \rho, \phi| l,p\rangle \langle l,p|~\rho d\rho ~d \phi.
\end{equation}
Next, using the inner product $\langle\rho,\phi|l_1,p_1\rangle = LG^{|l_1|}_{p_1}(\rho)e^{-il_1\phi}$,and  performing the integration over $\phi$ we get 
\begin{equation}
\hat{M} = \sum_{l,p}\sum_{l', p'} |l',p'\rangle \langle l,p| \int^{\infty} _{\rho=0} \left[LG^{|l'|}_{p'} (\rho)\right]^{*}LG^{|l|}_p (\rho) e^{i l' \pi} \delta_{l',-l} ~\rho d\rho ,
\end{equation} 
where $\delta_{l',-l}$ is the Kronecker delta. Summing over $l'$ and using the identity  \\ $\int^{\infty} _{\rho=0} \left[LG^{|l|}_{p'} (\rho)\right]^{*}LG^{|l|}_p (\rho) ~\rho d\rho = \delta_{p',p}$, we obtain
\begin{equation}
\hat{M}= \sum_{l,p}\sum_{ p'} |-l,p'\rangle \langle l,p| e^{-il \pi} \delta_{p', p}. 
\end{equation}
Finally, summing over $p'$, we obtain the following expression for the projector: 
\begin{equation}
\hat{M}= \sum_{l,p} e^{-il \pi}|-l,p\rangle \langle l,p| .
\end{equation}
We find that a reflection transforms the mode $| l,p\rangle$ into mode $| -l,p\rangle e^{-il \pi}$. Using the form of $\hat{M}$ above, we can show in a straightforward manner that $\hat{M}^{2n+1}=\hat{M}$ and $\hat{M}^{2n}=\mathbb{I}$, where $n$ is an integer. In other words, an even number of mirrors reflections transforms just as an identity operator whereas an odd number of reflections is equivalent to a single reflection.

\

\section{Mode transformation due to an image rotator}

As shown in Fig.~\ref{fig: ir_transformation}, an image rotator (IR) when kept at angle $\theta$, transforms the azimuthal phase profile $e^{-il\phi}$ into $e^{-il(\pi+2\theta-\phi)}$. Thus, an image rotator effectively transforms the state $|\rho, \phi\rangle $ into $|\rho, \pi +2\theta - \phi\rangle $. This transformation by an IR can be represented as $\hat{\rm IR} (\theta) |\rho, \phi\rangle = |\rho, \pi+ 2\theta - \phi\rangle$, where   $\hat{\rm IR}(\theta)$ is the projection operator. Using the similar procedure as worked out in Section~\ref{sec:mirror_operator}, we can write $\hat{\rm IR}(\theta)$  as:
\begin{figure}[t!]
\centering
\includegraphics[scale=1]{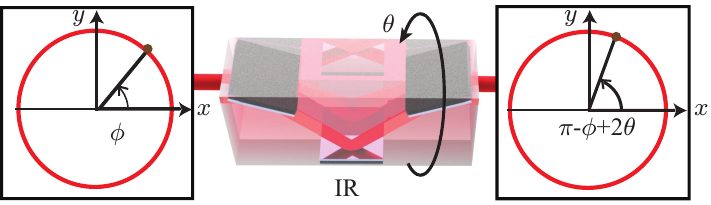}
\caption{{\bf Mode transformation due to the rotation caused by an image rotator(IR) orientated at angle $\theta$.} An incoming field with the phase profile $e^{il\phi}$ gets transformed into a field having the phase profile $e^{il(\pi -\phi + 2\theta)}$.}
\label{fig: ir_transformation}
\end{figure}
\begin{equation}
\hat{\rm IR}(\theta) =\int^{\infty} _{\rho=0}\int^{2 \pi}_{\phi=0} |\rho, \pi+ 2 \theta - \phi\rangle \langle \rho, \phi| ~\rho d\rho ~d \phi.
\end{equation}
Now, using the completeness condition $\sum_{l,p} |l,p\rangle \langle l,p| = \mathbb{I}$, we write $\hat{\rm IR}(\theta)$ in the LG basis as
\begin{equation}
\hat{\rm IR}(\theta) = \sum_{l,p}\sum_{l', p'}\int^{\infty} _{\rho=0}\int^{2 \pi}_{\phi=0} |l',p'\rangle \langle l',p'| \rho, \pi + 2\theta - \phi\rangle \langle \rho, \phi| l,p\rangle \langle l,p|~\rho d\rho ~d \phi.
\end{equation}
Next, using the inner product $\langle\rho,\phi|l_1,p_1\rangle = LG^{|l_1|}_{p_1}(\rho)e^{-il_1\phi}$, and  performing the integration over $\phi$ we get 
\begin{equation}
\hat{\rm IR}(\theta) = \sum_{l,p}\sum_{l', p'} |l',p'\rangle \langle l,p| \int^{\infty} _{\rho=0} \left[LG^{|l'|}_{p'} (\rho)\right]^{*}LG^{|l|}_p (\rho) e^{i l'\left( \pi+ 2\theta\right)} \delta_{l',-l} ~\rho d\rho .
\end{equation} 
where $\delta_{l',-l}$ is the Kronecker delta. Finally, summing over $l'$, using the identity \\  $\int^{\infty} _{\rho=0} \left[LG^{|l|}_{p'} (\rho)\right]^{*}LG^{|l|}_p (\rho) ~\rho d\rho = \delta_{p',p}$, and then summing over $p'$, we obtain 
\begin{equation}
\hat{\rm IR}(\theta) = \sum_{l,p} e^{-il\left( \pi+ 2\theta\right)} |-l,p\rangle \langle l,p|.
\end{equation}
Thus, an image rotator transforms the mode $| l,p\rangle$ into mode $| -l,p\rangle e^{-il (\pi+2\theta)}$.

\

\section{Derivation of the difference probability}

We consider the interferometric setup shown in Fig.~1C of the main text. The input state in the OAM basis is given by the density matrix:
\begin{align}\label{input_state}
\rho_{\rm in}= \sum^{+N}_{l_1,l_2=-N}\sum^{\infty}_{p_1,p_2 =0}  C^{p_1,p_2}_{l_1,l_2} |l_1, p_1\rangle\langle l_2,p_2|.
\end{align}
In this interferometer, there are two alternative ways in which an input mode $| l,p\rangle$ can reach the EMCCD plane. One is through the upper arm of the interferometer and second is through the lower arm containing the image rotator. We take the input field to be  $\hat{\bm x}$ polarized. In going through the upper arm, an input mode $| l,p\rangle$ faces an odd number of reflections and therefore transforms as
\begin{align}\label{upper-arm}
| l,p\rangle \hat{\bm x} \rightarrow |-l,p\rangle k_1 e^{-il\pi} e^{i(\beta_1-\omega_0t_1)} \hat{\bm x}.
\end{align}  
In going through the lower arm, the mode faces an even number of reflections and transmits through an image rotator kept at angle $\theta$, therefore, it transforms as
\begin{align}\label{lower-arm}
| l,p\rangle \hat{\bm x} \rightarrow |-l,p\rangle k_2 e^{-il(\pi+2\theta)} e^{i(\beta_2-\omega_0t_2)} \hat{\bm\epsilon}(\theta).
\end{align}  
Here, $\omega_0$ is the central frequency, and $k_1$ and $k_2$ are the  fractions of the incoming field passing through the two arms of the interferometer that depend on the reflection and the transmission coefficients of the beam splitters and mirrors; $t_1$ and $t_2$ are the photon travel times through the two arms, and $\beta_1$ and $\beta_2$ are the non-dynamical phases in the two arms. We note that when a mode with the state of polarization $\hat{\bm x}$ passes through an IR rotated at $\theta$, the state of polarization $\hat{\bm\epsilon}(\theta)$ of the transmitted field is given by \cite{karan2022ao}.%\cite{karan2022ao}. 
\begin{equation}
\hat{\bm \epsilon}\left( \theta\right)=  \cos\left[ \psi\left(\theta\right)\right] \hat{\bm x} +  \sin\left[ \psi\left(\theta\right)\right]e^{i\chi\left(\theta\right)} \hat{\bm y},
\end{equation}
where $\psi(\theta)$ and $\chi(\theta)$ represent the azimuth and ellipticity of the transmitted field. Using Eqs.~(\ref{upper-arm}) and (\ref{lower-arm}), we find that an input mode $| l,p\rangle$ in its passage through the interferometer gets transformed as
\begin{align}
| l,p\rangle \hat{\bm x} \rightarrow |-l,p\rangle k_1 e^{-il\pi} e^{i(\beta_1-\omega_0t_1)} \hat{\bm x}+|-l,p\rangle k_2 e^{-il(\pi+2\theta)} e^{i(\beta_2-\omega_0t_2)} \hat{\bm\epsilon}(\theta).
\end{align}
This transformation by the interferometer can be represented by the projection operator $\hat{P}$:
\begin{equation}
\hat{P}=\sum_{l=-\infty}^{\infty}\sum_{p=0}^{\infty}e^{i(-l\pi + \beta_1 -\omega_0 t_1 + \gamma_1)}
 \left[ |k_1|\hat{\bm x} +|k_2| e^{-i\left(\delta+2 l \theta\right)} ~\hat{{\bm\epsilon}}\left(\theta\right) \right]|-l, p\rangle\langle l, p|, \label{int-project}
\end{equation}
where we have substituted $k_1= |k_1|e^{i\gamma_1}$, $k_2 = |k_2|e^{i\gamma_2}$ and $\delta= (\beta_1 -\beta_2) - \omega_0(t_1 -t_2)+ \gamma_1- \gamma_2 $. In order to modify $\delta$, the upper arm of the interferometer has a geometric phase unit comprising of a quarter-wave plate (Q) with its first axis at $45^{\circ}$, a half-wave plate (H) with its first axis at $\frac{\delta}{2}$, and an additional quarter-wave plate with its fast axis at $45^{\circ}$.  In terms of the projection operator $\hat{P}$, the output state $\rho_{\rm out}$ corresponding to the input state $\rho_{\rm in}$ in Eq.~(\ref{input_state}) is given by
\begin{align}
\rho_{\rm out}= \hat{P}\rho_{\rm in}\hat{P}^{\dagger}.\label{output-state}
\end{align}
Therefore, the total single-photon detection probability  $I^{\delta}_{\rm out}(\theta)$ at the EMCCD plane as a function of $\theta$ is given by
\begin{equation}\label{out_intensity}
I^{\delta}_{\rm out}(\theta)=\int_{0}^{\infty}\int_{0}^{2\pi} \langle\rho, \phi | \rho_{\rm out} |\rho,\phi \rangle~\rho d\rho~ d\phi =\int_{0}^{\infty}\int_{0}^{2\pi} \langle\rho, \phi | \hat{P}\rho_{\rm in}\hat{P}^{\dagger} |\rho,\phi \rangle~\rho d\rho~ d\phi. 
\end{equation}
Using Eqs.~(\ref{input_state}), (\ref{int-project}) and (\ref{output-state}) and applying the vector relations $\hat{\bm x} \cdot \hat{\bm x}=\hat{\bm y} \cdot \hat{\bm y}=1$ and $\hat{\bm x} \cdot \hat{\bm y}=0$, we express $I^{\delta}_{\rm out}(\theta)$ as 
\begin{align}
&I^{\delta}_{\rm out}(\theta)= \sum_{l_1, l_2=-N}^{N}\sum_{p_1, p_2=0}^{\infty} C^{p_1,p_2}_{l_1, l_2}e^{i\left(l_2-l_1\right)\pi}\left[|k_1|^2 + |k_2|^2 e^{i 2\left(l_2 -l_1\right)\theta} \right. \notag \\
&\left. + |k_1||k_2|\cos\left[\psi\left(\theta\right)\right]\left\lbrace e^{i\left(\delta + 2l_2 \theta\right)}+ e^{-i\left(\delta + 2l_1 \theta\right)}\right \rbrace\right]  \int_{0}^{\infty}\int_{0}^{2\pi}\langle\rho,\phi|-l_1,p_1\rangle \langle-l_2,p_2|\rho,\phi\rangle \rho d\rho ~d\phi .
\end{align}
Now, using the identity $\int_{0}^{\infty}\int_{0}^{2 \pi}\left[LG_{p_1}^{|l_1|}(\rho)e^{-i l_1 \phi}\right]^{*}\left[LG_{p_2}^{|l_2|}(\rho)e^{-i l_2 \phi}\right]\rho d\rho~ d\phi = \delta_{l_1, l_2}\delta_{p_1, p_2}$, we write $I^{\delta}_{\rm out}(\theta)$ as 
\begin{equation}
I^{\delta}_{\rm out}(\theta)= \sum_{l=-N}^{N}\sum_{p=0}^{\infty} C^{p,p}_{l, l}\left[|k_1|^2 + |k_2|^2 + 2|k_1||k_2| \cos\left[\psi\left(\theta\right)\right] \cos\left(\delta + 2 l\theta\right)\right],
\end{equation}
The spectrum $S_l$ of an OAM mode with index $l$ is the detection probabilities summed over all the radial modes $p$ having the same OAM index $l$, that is, $S_l= \sum^{\infty}_{p=0}C^{p,p}_{l,l}$.  Thus we write $I^{\delta}_{\rm out}(\theta)$ as
\begin{equation}\label{angle-av intensity}
I^{\delta}_{\rm out}(\theta)= |k_1|^2 + |k_2|^2 + 2 |k_1||k_2|  \cos\left[\psi\left(\theta\right)\right]  \sum_{l=-N}^{+N} S_l \cos(\delta + 2 l\theta). 
\end{equation}
The normalization of the spectrum implies $\sum^{+N}_{l=-N} S_l =1$. Equation (\ref{angle-av intensity}) depicts the total probability at the output of the interferometer due to the input state $\rho_{\rm in}$. Nonetheless, in realistic experimental situations, the measured probability always contain some noise component  $I_n^\delta(\theta)$. In our experiments, the main source of noise is from the EMCCD detection. This is due to the fact that of the entire EMCCED camera pixels, only a small portion are involved in detection. These camera pixels are the ones that contribute to the signal. The remaining pixels contribute mostly to the noise because of the inherent dark counts of the EMCCD pixels. The other source of noise include ambient light, etc. Therefore, the measured probability $\bar{I}_{\rm out}^\delta(\theta)$ at the camera plane can be written as  
\begin{equation} \label{out_inten_noise}
\bar{I}_{\rm out}^\delta(\theta)= I_n^\delta(\theta)+|k_1|^2 + |k_2|^2  + 2 |k_1||k_2|\cos\left[\psi\left(\theta\right)\right] \sum_{l=-N}^{+N}  S_l \cos(\delta + 2 l\theta).
\end{equation}
In order to bypass the effects due to noise sources, we use  the two-shot technique \cite{kulkarni2017natcomm}. We take two probability measurements, one at $\delta= \delta_{c}$ and the other at $\delta=\delta_{d}$.  Assuming that the shot-to-shot noise remains constant, that is,  $ I_n^{\delta_c}(\theta)  \approx I_n^{\delta_d}(\theta)$, we write the difference probability $\Delta\bar{I}_{\rm out}(\theta) = \bar{I}_{\rm out}^{\delta_c}(\theta) - \bar{I}_{\rm out}^{\delta_d}(\theta)$ as %\cite{kulkarni2017natcomm}
\begin{equation}\label{deltaI_sl}
\Delta\bar{I}_{\rm out}(\theta)= 2 |k_1||k_2|\cos\left[\psi\left(\theta\right)\right]  \sum_{l=-N}^{+N}   S_l [ \left(\cos \delta_c - \cos \delta_d \right)  \cos {2 l \theta} -\left(\sin \delta_c - \sin\delta_d \right)  \sin {2 l \theta}] .
\end{equation}
We assume that the spectrum is symmetric, that is,  $S_l=S_{-l}$. Therefore $\cos(-2l\theta)=\cos(2l\theta)$, and $\sin(-2l\theta)=-\sin(2l\theta)$. The difference probability $\Delta\bar{I}_{\rm out}(\theta)$  can now be written as 
\begin{equation}\label{deltaI_sl-symm}
\Delta\bar{I}_{\rm out}(\theta)= 2 |k_1||k_2|\cos\left[\psi\left(\theta\right)\right]\sum_{l=-N}^{N}  S_l (\cos \delta_c - \cos\delta_d) \cos{2l\theta} .
\end{equation}
We note that $\Delta\bar{I}_{\rm out}(\theta)$  contains the term $\cos\left[\psi\left(\theta\right)\right]$. This is because of the fact that an IR rotates the polarization of incident field. Although it would be ideal to have an image rotator which does not induce any $\theta$-dependent polarization changes, no such image rotator exists currently. Nonetheless, we can bypass this polarization effect by redefining the difference probability as the polarization-corrected difference probability $\Delta I_{\rm out}(\theta)= \Delta\bar{I}_{\rm out}(\theta)/\cos\left[\psi\left(\theta\right)\right]$ such that  
\begin{equation}\label{deltaI_pol_corrected_si}
\Delta I_{\rm out}(\theta)= 2 |k_1||k_2| \sum_{l=-N}^{N}  S_l (\cos \delta_c - \cos\delta_d) \cos{2l\theta}.
\end{equation}
We can measure $\cos\left[\psi\left(\theta\right)\right]$ in the following manner. We note that $\cos\left[\psi\left(\theta\right)\right]$ is the projection of the polarization vector along $\hat{\bm x}$. Therefore, $|\cos\left[\psi\left(\theta\right)\right]|^2$ is the ratio of the probability $I_{2x}(\theta)$ along $\hat{\bm x}$-direction and the total probability $I_{2}^{\rm tot}$, that is,
\begin{align}
|\cos\left[\psi\left(\theta\right)\right]|^2=\frac{|\hat{\bf \epsilon}(\theta)\cdot\hat{\bf x}|^2}{|\hat{\bf \epsilon}(\theta)|^2}=\frac{I_{2x}(\theta)}{I_2^{\rm tot}}.
\end{align}
Therefore, $\cos\left[\psi\left(\theta\right)\right]=\sqrt{\frac{I_{2x}(\theta)}{I_2^{\rm tot}}}$. We note that $I_2^{\rm tot}$ is independent of $\theta$ because although an IR rotates the wavefront and polarization, it does not change the probability of the field passing through it. The probabilities $I_2^{\rm tot}$ and $I_{2x}(\theta)$ can be measured by blocking the upper arm of the interferometer. The probability through the IR in this case is $I_2^{\rm tot}$. For measuring  $I_{2x}(\theta)$, one needs to put a polarizer oriented along $\hat{\bm x}$ direction right after the IR and then measure the probability as a function of $\theta$. The polarization corrected difference probability $\Delta I_{\rm out}(\theta)$ is used for obtaining the OAM spectrum of the input state $\rho_{\rm in}$.

\section{Measurement of non-symmetric OAM spectrum}

In this section, we work out how our technique works for an input state $\rho_{\rm in}$ with non-symmetric OAM spectrum, that is, in situation in which the OAM spectrum is not centered around $l=0$, and $S_l\neq S_{-l}$. Since the assumption $S_l=S_{-l}$ is no longer valid, one cannot measure the spectrum using a two-shot technique. However, we show that it is indeed possible to measure a  non-symmetric spectrum in our scheme using a four-shot technique. Unlike in the case of a symmetric spectrum, in which one requires only two intensity measurements at $\delta=0$ and $\delta=\pi$, a non-symmetric spectrum requires four intensity measurements, one each at  $\delta=0$, $\delta=\pi$, $\delta=\pi/2$ and $\delta=3\pi/2$. Using equation~(\ref{deltaI_sl}), we obtain the following expression for the difference probability $\Delta \bar I^{(0,\pi)}_{\rm out}\left(\theta \right)$, for $\delta_c=0 $ and $\delta_d=\pi$: 
\begin{equation}\label{deltaI_sl_0_pi}
\Delta \bar I^{(0,\pi)}_{\rm out}\left(\theta \right)= 4 |k_1||k_2|\cos\left[\psi\left(\theta\right)\right]  \sum_{l=-N}^{+N}   S_l   \cos {2 l \theta}.
\end{equation}
Similarly, taking $\delta_c=3\pi/2 $ and $\delta_d=\pi/2$, and using Eq.~(\ref{deltaI_sl}), we obtain the following expression for the difference probability $\Delta \bar I^{(3\pi/2,\pi/2)}_{\rm out}\left(\theta \right)$:
\begin{equation}\label{deltaI_sl_3pi2_pi2}
\Delta \bar I^{(3\pi/2,\pi/2)}_{\rm out}\left(\theta \right)= 4 |k_1||k_2|\cos\left[\psi\left(\theta\right)\right]  \sum_{l=-N}^{+N} S_l \sin {2l\theta}.
\end{equation}
Just as in the case of the symmetric spectrum, in order to bypass the effects due to polarization, we work with the polarization corrected difference probabilities $\Delta I^{(0,\pi)}_{\rm out}\left(\theta \right)=\Delta \bar I^{(0,\pi)}_{\rm out}\left(\theta \right)/\cos\left[\psi\left(\theta\right)\right] $, and $\Delta I^{(3\pi/2,\pi/2)}_{\rm out}= \Delta \bar I^{(3\pi/2,\pi/2)}_{\rm out}/ \cos\left[\psi\left(\theta\right)\right] $. Therefore,
\begin{align*}
& \Delta I^{(0,\pi)}_{\rm out}\left(\theta \right)=4 |k_1||k_2| \sum_{l=-N}^{+N}   S_l   \cos {2 l \theta} \\
& \Delta I^{(3\pi/2,\pi/2)}_{\rm out}(\theta)= 4 |k_1||k_2|  \sum_{l=-N}^{+N}   S_l   \sin {2 l \theta}
\end{align*}

Next, we define the measured OAM spectrum $\bar S_l$ as 
\begin{align}\label{deltaI_four_shot}
\bar S_l &\equiv \int^{\pi}_{0}\left[\Delta I^{(0,\pi)}_{\rm out}\left(\theta \right) + i \Delta I^{(3\pi/2,\pi/2)}_{\rm out}\left(\theta \right)\right]e^{-i l 2\theta} d\theta \nonumber \\  
&= 4 \pi|k_1||k_2| \sum_{l'=-N}^{+N}   S_{l'}\delta_{l,l'}\\
&= 4 \pi|k_1||k_2|S_l.
\end{align}
Although the measured spectrum $\bar S_l$ is not normalized, we find that it is proportional to the true spectrum $S_l$. Now using the fact that ${\rm Tr}\left(\rho_{\rm in}\right)=1$, we obtain the true OAM spectrum $S_l$ as
\begin{equation}
S_l= \dfrac{\bar S_l}{\sum^{+N}_{-N}\bar S_l}.
\end{equation}
Thus,  one can measure any OAM spectrum $S_l$ using four intensity measurements. In situation in which one has the prior information that the spectrum is symmetric, one can measure the spectrum using only two intensity measurements. Otherwise, using four intensity measurement, one can measure any spectra, symmetric or asymmetric.   

\section{Effects due to the angular deviation of the image rotator}
\begin{figure}[htbp]
\centering
\includegraphics[scale=0.92]{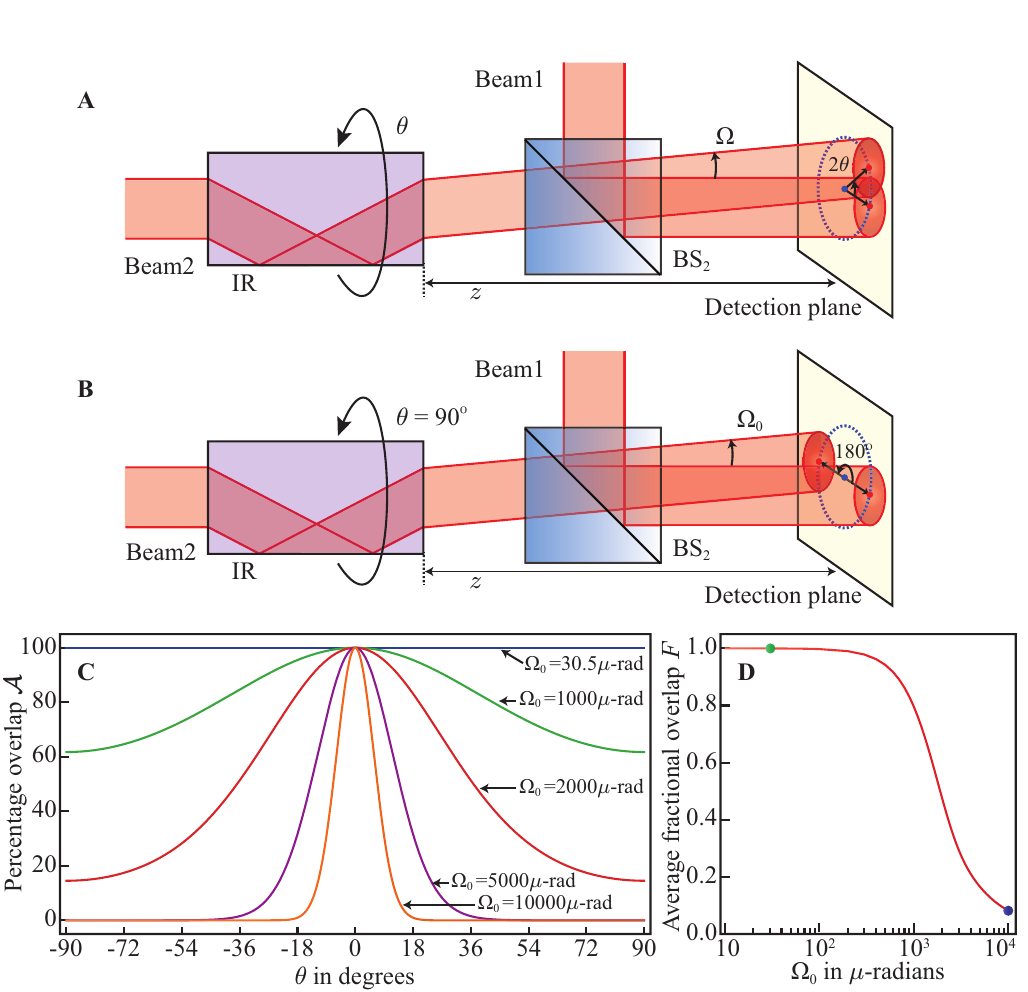}
\caption{{\bf Conceptual illustration and numerical analysis of the effect of angular deviation of image rotator.} {\bf(A)} Schematic representation of the overlap of Beam1 and Beam2 at the detection plane at a distance $z$ from the image rotator (IR). The angular deviation of Beam2 with respect to Beam1 is defined as $\Omega$. The dotted blue circle traces the trajectory of the center of Beam2 as a function of $\theta$. {\bf(B)} Depiction of the angular deviation at $\theta= 90^{\circ}$, at which the angular deviation reaches its maximum value $\Omega_0$. {\bf(C)} Plot of percentage overlap $\mathcal{A}$ as a function of $\theta$ at five different $\Omega_0$ values. {\bf(D)} The plot of average fractional overlap ($F$) as a function of $\Omega_0$. $\Omega_0 = 30.5~\mu$-radians (green dot) and $\Omega_0 = 10000~\mu$-radians (blue dot) correspond to the home-built IR and a typical commercial Dove prism, respectively.}
\label{fig:conceptual_ang_deviation}
\end{figure}

The accuracy of reconstructing the OAM spectrum  through the experimental setup of Fig.~1C of the main text entirely depends on how accurately one is able to measure $\bar{I}_{\rm out}^\delta(\theta)$ of equation~(\ref{out_inten_noise}) as a function of rotation angle $\theta$. Since $\bar{I}_{\rm out}^\delta(\theta)$ is the intensity due to the interference of the two fields coming through the second beam splitter (BS$_2$) of Fig.~1C, it is important that the two fields have perfect overlap at the output of the beam splitter (BS$_2$). Any deviation from the perfect overlap results in error in the measurement of $\bar{I}_{\rm out}^\delta(\theta)$ and thus in the estimated spectrum. However, due to manufacturing issues, IRs invariably introduce angular deviations in the transmitted field and cause the overlap between the two fields to decrease as a fucntion of $\theta$. This is depicted in Figure~\ref{fig:conceptual_ang_deviation}. The field reflecting from (BS$_2$) is referred to as Beam1 and that transmitting through it is referred to as Beam2. The detection plane is located at distance $z$ from the IR. When $\theta=0$, both beams are aligned perfectly with complete overlap at the detection plane. However, as $\theta$ increases, rotation of IR causes angular deviation of Beam2 resulting in imperfect overlap as illustrated in Fig.~\ref{fig:conceptual_ang_deviation}A. At $\theta=90^\circ$, IR introduces maximum angular deviation, represented as $\Omega_0$, resulting in the worst overlap, as shown in Fig.~\ref{fig:conceptual_ang_deviation}B.

In order to numerically analyze this overlap as a function of $\theta$, we first define the center ($x_{c2}$ and $y_{c2}$) of Beam2 as
\begin{align}
&x_{c2} = z \tan \Omega \left[\cos 2\theta -1 \right],\label{eq:xc2}\\
&y_{c2} = z \tan \Omega \sin 2\theta. \label{eq:yc2}
\end{align}
We note that at $\theta =0^{\circ}$, $x_{c2}=0$ and $y_{c2} =0$, which implies that Beam1 and Beam2 are perfectly aligned. On the other hand, at $\theta=90^{\circ}$, $x_{c2}=- 2 z \tan \Omega_0$ and $y_{c2} =0$, and this corresponds to the maximum angular deviation $\Omega_0$. Next, we represent the electric fields of Beam1 and Beam2 at transverse position $(x,y)$ on the detection plane by $E_1\left(x-x_{c1}, y- y_{c1}\right)$ and  $E_2 \left(x - x_{c2},y - y_{c2}\right)$, respectively, and define the percentage overlap  $\mathcal{A}$ of the two fields as 
\begin{align}
\mathcal{A} = \dfrac{|\iint E_1 \left(x - x_{c1},y-y_{c1}\right) E^{*}_2\left(x-x_{c2}, y- y_{c2}\right) dx dy |^2}{\iint |E_1 \left(x - x_{c1},y-y_{c1}\right)|^2 dx dy   \iint |E_2\left(x-x_{c2}, y- y_{c2}\right)|^2 dx dy } \times 100 \%.
\end{align}
We numerically evaluate $\mathcal{A}$ as a function of $\theta$ for five different $\Omega_0$ and plot them in Fig.~\ref{fig:conceptual_ang_deviation}C. The distance to the detection plane is taken to be $z = 250$mm. We find that, for $\Omega_0 = 10000\mu$-radians, which is typically observed with commercial Dove prisms,  the overlap $\mathcal{A}$ becomes zero within an angular rotation of about $\pm 18^\circ$. However, for $\Omega_0 = 30\mu$-radians, which is what is achieved with our home-built IR, the overlap $\mathcal{A}$ remains almost constant over the entire range of $\theta$. The percentage overlap $\mathcal{A}$ over the entire range decides how accurately the interferometer is able to measure $\bar{I}_{\rm out}^\delta(\theta)$. We quantify this in terms of average fractional overlap ($F$) under the $\mathcal{A}$ versus $\theta$ curve. The fraction is defined by taking $F=1$ at $\Omega_0 =0$. We numerically evaluate $F$ as a function of $\Omega_0$ and plot it in Fig.~\ref{fig:conceptual_ang_deviation}D. We note that at $\Omega_0 = 10000\mu$-radians, represented by the blue dot, that is typically observed with commercial Dove prism, the average fractional overlap $F=0.0817$. This clearly shows why commercial Dove prism cannot yield accurate estimation of $\bar{I}_{\rm out}^\delta(\theta)$ and thus of the OAM spectrum. On the other hand, at $\Omega_0 = 30\mu$-radians, which is shown by the green dot and which corresponds to our home-built IR, the average fractional overlap $F=0.9998\approx 1$, implying close to perfect estimation of $\bar{I}_{\rm out}^\delta(\theta)$ resulting in near perfect reconstruction fidelity of the OAM spectrum reported in our experiments. 

We note that the above analysis is for a detection scheme that assumes no prior knowledge of the input OAM spectrum. However, with prior knowledge, the inaccuracies caused by the angular deviation of image rotators could be bypassed to the extent that one has the prior information.

\end{document}